\documentclass[twocolumn,superscriptaddress,prb]{revtex4-1}

\usepackage{amsmath}
\usepackage{amsfonts}
\usepackage{color}
\usepackage[colorlinks,bookmarks=false,citecolor=blue,linkcolor=red,urlcolor=blue]{hyperref}
\usepackage{ulem}
\setcitestyle{numbers,square}
\usepackage{graphicx,subfigure}

\def \be {\begin{equation}}
\def \ee {\end{equation}}
\def \ba {\begin{array}}
\def \ea {\end{array}}
\def \bea {\begin{eqnarray}}
\def \eea {\end{eqnarray}}

\def \ble {\begin{widetext}\begin{equation}}
\def \ele {\end{equation}\end{widetext}}
\def \blea {\begin{widetext}\begin{eqnarray}}
\def \elea {\end{eqnarray}\end{widetext}}

\def \r {\rho}

\def \f {\frac}

\def \ep {\mathrm{e}}

\def \tr {\mathrm{tr}}

\def \and {{\mathrm{and}}}

\begin{document}


\title{Universal logarithmic correction to R\'enyi (Shannon) entropy in generic systems of critical quadratic fermions }

\author{Babak Tarighi}

\affiliation{Department of Physics, University of Mohaghegh Ardabili, P.O. Box 179, Ardabil, Iran}

\author{Reyhaneh Khasseh}

\affiliation{Theoretical Physics III, Center for Electronic Correlations and Magnetism,
Institute of Physics, University of Augsburg, D-86135 Augsburg, Germany}
\affiliation{Max-Planck-Institut f\"ur Physik Komplexer Systeme,N\"othnitzer Strasse 38, D-01187, Dresden, Germany}

\author{M. N. Najafi}

\affiliation{Department of Physics, University of Mohaghegh Ardabili, P.O. Box 179, Ardabil, Iran}
\author{M.~A.~Rajabpour}
\affiliation{Instituto de Fisica, Universidade Federal Fluminense,\\
Av.~Gal.~Milton Tavares de Souza s/n, Gragoat\'a, 24210-346, Niter\'oi, RJ, Brazil}

\begin{abstract}
The R\'enyi (Shannon) entropy, i.e. $Re_{\alpha}(Sh)$, of the ground state of quantum systems in local bases normally show a volume-law behavior. For a subsystem of quantum chains at critical point there is an extra logarithmic subleading term with a coefficient which is universal. In this paper we study this coefficient for generic time-reversal translational invariant quadratic critical free fermions.
These models can be parameterized by a complex function which has zeros on the unit circle. When the zeros on the unit circle do not have degeneracy and there is no zero outside of the unit circle we are able to classify the coefficient of the logarithm.
In particular, we numerically calculate the R\'enyi (Shannon) entropy in configuration basis for wide variety of these models and show that there are two distinct classes.
For systems with $U(1)$ symmetry the coefficient is proportional to the central charge, i.e. one half of the number of points that one can linearize the dispersion relation of the system; for all the values of $\alpha$ with transition point at $\alpha=4$. For systems without this symmetry, when $\alpha>1$ this coefficient is again proportional to the central charge. However, the coefficient for $\alpha\leq 1$
is a new universal number.  
Finally, by using the discrete version of Bisognano-Wichmann modular Hamiltonian of the Ising chain we show that these coefficients are universal and dependent on the underlying CFT.

\end{abstract}

\maketitle





\section{Introduction}

In quantum mechanics the outcome of a measurement of an observable is one of the eigenvalues of the observable. Each outcome happens with a particular probability. These probabilities can be used to calculate R\'enyi (Shannon) entropy which is a representative number for the probability distribution. The number depends on the chosen observable and gives an idea about the distribution of probabilities. For many-body systems there are many possibilities to choose the observable and study its distribution and extract interesting information. In quantum chains one can look to a local observable defined on each site and find the probability of having a particular configuration for the full system in, for example, the ground state. This will lead to a set of probabilities that its size grows linearly with the size of the Hilbert space. In quantum spin chains, when one takes the ground state, the  R\'enyi (Shannon) entropy in $\sigma^{x,y,z}$ basis present some information about the phase transition and the universality class\cite{Wolf:2008,Jean-Marie:2009,Jean-Marie:2010,Luitz:2014,Luitz:2014B,Luitz:2014C}. Instead of calculating the R\'enyi (Shannon) entropy of the full system one can use marginal probabilities and calculate the same quantities for the subsystem. These quantities as their full system counterparts also show a volume law behavior, however, at the phase transition point there is a logarithmic subleading term
the coefficient of which shows interesting {\it{universal}} behavior 
\cite{AR:2013,Jean-Marie:2013,Lau:2013,Stephan:2014,AR:2014,AR:2015,NR2016,Alcaraz:2016,Getelina:2016}. Studies on many different quantum critical spin chains reveal that the coefficient of the logarithm depends on the chosen basis but shows some level of universality in some particular bases dubbed as conformal basis \cite{AR:2014}. In these models there are infinite possibilities to choose the local observable and it seems any kind of classification is hopeless. In fermionic systems the situation seems more tractable. The most obvious local observable to take is the number operator. One can write the ground state in configuration basis and look to the probabilities of different configurations. These probabilities are dubbed as formation probabilities and have been studied for subsystems of certain free fermions in depth \cite{Franchini:2005,NR2016,NR:2019,Ares:2020,ARV:2021}. For results on the full system see \cite{Jean-Marie:2010,ARV:2020}. These probabilities have been also investigated in experiments \cite{Zhang:2017}.

Time-reversal translational invariant quadratic critical free fermions show interesting phase transitions. Depending on the couplings one can produce critical systems with integer and half integer central charges \cite{Its:2008,Verresen:2019}. They also show interesting topologically protected phases \cite{Verresen:2018,Verresen:2019}. In addition there are many efficient methods  to calculate the formation probabilities for extremely large systems \cite{NR2016,NR:2019}. These methods are also useful to work directly with subsystems embedded in the systems with infinite size. This is very useful to avoid the problem of finite size effect regarding the full system. We notice that since the number of probabilities grows exponentially with the size of the subsystem there is an unavoidable limitation on the size of the subsystem that one can take in numerical calculations. In this paper we make a step in full classification of the the coefficient of the logarithmic term in the R\'enyi (Shannon) entropy of generic time-reversal translational invariant quadratic critical free fermions. We calculate this quantity for various critical models and show that the coefficient is proportional to the number of points that one can linearlize the dispersion relation but the proportionality constant is very much dependent on the presence (absence) of the $U(1)$ symmetry. In systems with $U(1)$ symmetry clear picture emerges for the coefficient of the logarithm with respect to $\alpha$. However, for systems without this symmetry the picture is clear 
just for $\alpha>1$.

The paper is organized as follows: In Sec. \ref{sec:Setup and definitions}
we first define the R\'enyi (Shannon) entropy for the subsystem. To extract the coefficient of the logarithm we define the quantity $I_{\alpha}$ for two subsystems of our original subsystem which was embedded in an infinite system. The setup used in this paper has not been considered previously. Most of the previous studies worked with a system which was periodic finite system and partitioned the system into two parts\cite{AR:2013,Stephan:2014,AR:2014}. In our setup we have a tri-partite situation. 

In Sec. \ref{sec:Models and Methods of calculation} we introduce the kind of models that we considered in this study, i.e.  time-reversal translational invariant quadratic critical free fermions. Apart from their physical appeal these models provide series of different universality classes. They can be solved exactly and one can find the desired formation probabilities exactly and efficiently in the thermodynamic limit. We categorize these models to two types, those with and without $U(1)$ symmetry. We also show how one can find the formation probabilities out of the correlation matrices for these models. A couple of interesting dualities regarding the correlation matrices of different models will be also presented in this section.

In Sec. \ref{sec:Summary of results} we summarize our main results. We make a few conjectures regarding the coefficient of the logarithm in the models that we considered. It seems there are two classes. Those that have $U(1)$ symmetry and models without manifest $U(1)$ symmetry. In the latter models we just consider models where the corresponding $f(z)$ function  does not have zero outside of the unit circle. 

In Sec. \ref{sec:Numerical and fitting procedure} we briefly describe our numerical and fitting procedure. Then in Sec. \ref{sec:Details of the analysis} we present the details of the models that we considered and provide support for the results presented in Sec. \ref{sec:Summary of results}. In Sec. \ref{sec:BW} we use the discrete version of Bisognano-Wichmann modular Hamiltonian for the Ising chain and show that the results converge rapidly to the exact results. 
Finally in Sec. \ref{sec:Discussion} we discuss the results further 
and then conclude the paper in Sec. \ref{sec:Conclusions}. 

The paper is accompanied with two appendices. In the first appendix \ref{App:fitting} we provide the details of the fitting methods that we have used to extract the coefficient of the logarithm. In the appendix \ref{App:Shannon} we provided the exact Shannon entropy of the models that we considered for different sizes.

\section{Setup and definitions}\label{sec:Setup and definitions}

In this section we present the basic definitions and the setup of the problem.
The quantities of interest, R\'enyi and Shannon entropies are defined as follows:
Consider the normalized ground state of a quantum chain
Hamiltonian, i.e. $|g\rangle=\sum_I a_I|I\rangle$, expressed in a particular local bases $|I\rangle = |i_1, i_2,...,i_N \rangle$, where $N$ is the system size and $i_1, i_2,...,i_N$ are the
eigenvalues of some local operators defined on the lattice
sites. The R\'enyi and Shannon entropies of the total system with size $N$ are defined as

\bea \label{Renyi-total}
Re_{\alpha}(N) &=& \f{1}{1-\alpha} \ln \sum_I P_I^{\alpha},\\
Sh(N)&=&-\sum_I   P_I\ln P_I,
\eea
where $P_I = |a_I|^2$ is the probability of finding the system in the particular configuration given by $|I\rangle$. These probabilities are dubbed as formation probabilities in \cite{NR2016}. In the above definition $\alpha$ can be any positive real number. Note that $\alpha\to 1$ gives us just the Shannon entropy.

\begin{figure}[htp]
  \centering
  \includegraphics[width=0.5\textwidth]{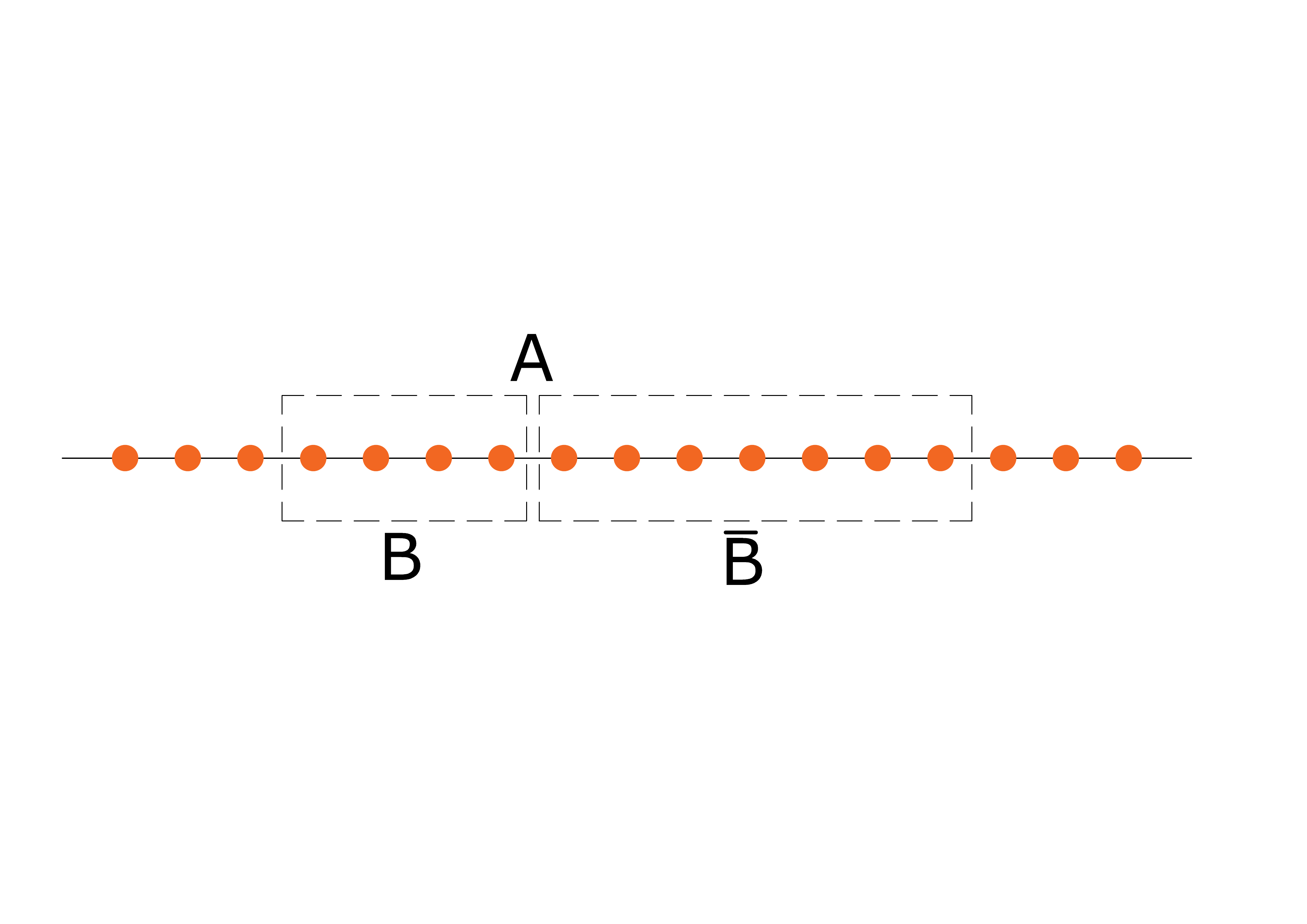}\\
  \caption{The setup used to calculate $I_{\alpha}(\ell)$. Here $B$ and $\bar{B}$ have sizes $\ell$ and $L-\ell$ respectively. }
  \label{setup}
\end{figure}

By considering local bases it is always possible to decompose the configurations as a combination
of the configurations inside and outside of a subregion $A$
as $|I\rangle = |I_AI_{\bar{A}}\rangle$, where $I_A$ and $I_{\bar{A}}$ are the sub-configurations corresponding to $A$ and $\bar{A}$. Then, one can define the marginal probabilities as $p_{I_A} =\sum_{I_{\bar{A}}}P_{I_AI_{\bar{A}}}$. Using these probabilities one can now define the R\'enyi and Shannon entropies of the subsystem with size $L$ as follows:

\bea \label{Renyi-subsystem}
Re_{\alpha}(L) &=& \f{1}{1-\alpha} \ln \sum_{I_A} p_{I_A}^{\alpha},\\
Sh(L)&=&-\sum_{I_A}   p_{I_A}\ln p_{I_A}.
\eea
The above two quantities at the critical point normally behave as 
\bea \label{Renyi-subsystem-critical-behavior}
Re_{\alpha}(L) &=& a_{\alpha}L+x_{\alpha}\ln L+\mathcal{O}(1),\\
Sh(L)&=&a_{1}L+x_{1}\ln L+\mathcal{O}(1).
\eea
The quantities of interest in this paper are $x_{\alpha}$ and $x_{1}$. To isolate these quantities one can divide the region $A$ to two subsystems $B$ and $\bar{B}$ with sizes $\ell$ and $L-\ell$ respectively, see Fig \ref{setup}\label{Fig}.
Then one can define
\bea \label{I}
I_{\alpha}(\ell) &=& Re_{\alpha}(\ell)+Re_{\alpha}(L-\ell)-Re_{\alpha}(L),\\
I_{1}(\ell)&=&Sh(\ell)+Sh(L-\ell)-Sh(L).
\eea
In the rest of paper, we consider the case $\ell=\f{L}{2}$. Then we expect
\bea \label{I-critical-behavior}
I_{\alpha}\Big{(}\f{L}{2}\Big{)} &=& x_{\alpha}\ln L+\mathcal{O}(1),\\
I_{1}\Big{(}\f{L}{2}\Big{)}&=&x_{1}\ln L+\mathcal{O}(1).
\eea
We calculate the above two quantities for different infinite size, i.e. $N\to\infty$, critical systems. The advantage of this setup is that we are free from the finite size effects of the total system and just bounded with the limitations coming from the subsystem size itself.

\section{Models and Methods of calculation}\label{sec:Models and Methods of calculation}

In this section we first define our Hamiltonian of interest and then present the formulas that one can use to calculate the formation probabilities and ultimately the $I_{\alpha}$. Here we follow the notation in \cite{Verresen:2019}

The Hamiltonian of the most general translational invariant (periodic) quadratic fermionic chain with time-reversal symmetry takes the form
\be \label{fermionicgeneric}
H = \sum_{r=-R}^{R}\sum_{j\in \Lambda}^N 
\Big[ A_r c_j^{\dagger}c_{j+r}
    + \frac{B_r}{2}( c_j^{\dagger}c_{j+r}^{\dagger}-  c_j c_{j+r})
\Big]
+\text{const},
\ee
with the local fermionic modes $c_j$, $c_j^\dag$ and the parameters $A_r=A_{N-r}$, $B_r=-B_{N-r}$ and $\Lambda$ represents the sites of the lattice. The above Hamiltonian can be exactly diagonalized  after going to the Fourier space and Bogoliubov transformation as follows:
\be \label{diagonalized form}
H = \sum_k|f(e^{ik})|\eta_k^{\dagger}\eta_k+\text{const},
\ee
where we have defined
\be \label{diagonalized form}
\eta_k=\f{1}{2}(1+\f{f(e^{ik})}{|f(e^{ik})|})c_k^{\dagger}+\f{1}{2}(1-\f{f(e^{ik})}{|f(e^{ik})|})c_{-k},
\ee
and the sum over $k$ goes over momenta $k_n=2\pi n/N$. All the information about the couplings are in the complex function $f(z)$ which we defined as 
\be \label{f(z) definition}
f(z):=\sum_{m} t_{m}z^{m},
\ee
where we have
\bea \label{t and couplings}
A_r=-\f {t_r+t_{-r}}{2},\\
B_r=-\f {t_r-t_{-r}}{2}.
\eea
In this system the vacuum is defined as $\eta_k|0\rangle=0$ for $\forall k$.
When $B_r\neq 0$ the vacuum state is the ground state, while when $B_r=0$ the Hamiltonian has  $U(1)$ symmetry which means the particle number is conserved. In this case one needs to fill the negative modes depending on the number of particles in the system to reach to the ground state.

\subsection{Formation probabilities}\label{subsec:Formation probabilities}

Before concentrating on critical  models explicitly, since all the forthcoming calculations are based on the correlation matrix we briefly define it here. The correlation matrix $\textbf{G}$ for the eigenstates is defined using two Majorana fermionic operators $\gamma_j\equiv c_j^{\dagger}+c_j$, and $\bar{\gamma}_j\equiv i\left( c_j^{\dagger}-c_j\right) $ as follows:
\begin{equation}
	iG_{jk}=\langle g|\bar{\gamma}_j\gamma_k|g\rangle.
	\label{Eq:CorrelationMatrix}
\end{equation}
One can use the above matrix to calculate all the observables in this system. For example, in this system the formation probabilities are defined as follows:
consider the ground state of the system written in configuration basis. That means each configuration of fermions can appear with particular probability in the ground state. These probabilities can be calculated using the following formula \cite{NR:2019,NR2016}

\bea \label{formation probability}
p(C)=\det (\f{\bf{I}-\bf{I}_c.\bf{G}}{2}),
\eea
where $\bf{I}$ is an identity matrix and $\bf{I}_c$ is a diagonal matrix
made out of $\pm1$. We set its diagonal element to $-1$ when we have a fermion and $+1$ when there is
no fermion at the corresponding site. When $\bf{I}_c=\bf{I}$ the corresponding probability is called emptiness formation probability.
The above formula works for the full(sub) system if one takes the $\textbf{G}$ matrix of full(sub) system. It also works for  disjoint intervals as far as one takes the $\textbf{G}$ matrix of the subsystem. Using the determinant properties it is easy to show that the set of formation probabilities is the same for the matrices $\bf{G}$, $-\bf{G}$, $\bf{G}^T$ and $-\bf{G}^T$. In other words, although the associated probabilities for different configurations might change, the whole set is the same.
Even more generally the matrices $\bf{I}_c.\bf{G}$, $\bf{G}.\bf{I}_c$ and $\bf{I}_c.\bf{G}.\bf{I}_c$ have the same set of formation probabilities and consequently the same Shannon and R\'enyi entropies.
To summarize, different models with different correlation matrices might have the same R\'enyi entropies.

\subsection{Critical systems}\label{subsec:Critical systems}

It is known that when the complex function $f(z)$ has zeros on the unit circle the ground state is critical and depending on the number of zeros one can have different universality classes with different central charges, for a review see \cite{Shankar:1994}. The reason behind this fact is that when $f(z)$ has zeros on the unit circle one can linearize the dispersion relation $|f(e^{ik})|$ around that momentum and get one gappless Majorana fermion. This Majorana fermion  contributes $c=\f{1}{2}$ to the central charge of the system so that we finally have $c=\f{N_l}{2}$, where $N_l$ is the number of zeros on the unit circle.

In general one can think about two types of critical systems, those with $U(1)$ symmetry and those without this symmetry. In this work, we show that the behavior of the $x_{\alpha}$ is very much dependent on the presence or absence of $U(1)$ symmetry. Because of that we will study these two cases separately.

\subsubsection{Models with $U(1)$ symmetry} \label{subsubsec:Models with $U(1)$ symmetry}

In these Hamiltonians we have $B_r=0$. A good examples of these types of Hamiltonians are the ones with the following $f(z)$ function:

\bea \label{U(1) models}
f_z(n)=-(z^n+z^{-n}),
\eea
which corresponds to $U(1)$-symmetric $n$-step hopping fermions.
It has the central charge $c=n$. The $n=1$ is the celebrated simple hopping chain.
For half filling case the correlation matrix of the ground state is

\bea \label{U(1) models}
\bf{G}&=&2\bf{C}-\bf{I},\\
C_{jk}&=&\f{1}{\pi(j-k)}\sum_{m=1}^n(-1)^{m+n}\sin[\f{\pi(2m-1)(j-k)}{2n}].\hspace{0.5cm}
\eea
The diagonal elements can be found by taking the limit. In principle, it is possible to consider more complicated models such as $f_z(\{a_n\})=\sum_n a_nf_z(n)$. The central charge is dependent again on the number of points where one can linearize the dispersion relation and very much dependent on the constants $a_n$. For example, consider the case $f_z(\{a_1,a_2\})=a_1f_z(1)+a_2f_z(2)$. For $|a_1|\geq |a_2|$ we have just two points to linearize the dispersion relation and we expect $c=1$, however, for $|a_2|> |a_1|$ we have four points to linearize so we expect $c=2$. The $\bf{C}$ matrix in this case can be written as
\bea \label{C matrix general}
C_{jk}(a_1,a_2)=\hspace{5.5cm}\nonumber\\
\f{1}{\pi(j-k)}( \sin [k_1^* (j-k)]- \sin [k_2^* (j-k)]);\hspace{0.75cm}
\eea
where $k_1^*\geq k_2^*$ are the solutions of the equation $a_1 \cos[k]+a_2 \cos[2k]=0$ in the range $(0,\pi)$. The diagonal elements can again be found by taking the limit.

In this work we will study $f_z(1)$, $f_z(2)$ and $f_z(\{a_1,a_2\})$ with $(a_1,a_2)=\{(1,1),(1,2)\}$,  and make a general statement about the behavior of $x_{\alpha}$. 
\\

\subsubsection{Models without $U(1)$ symmetry} \label{subsubsec:Models without $U(1)$ symmetry}

A fairly general form of $f(z)$ with zeros on the unit circle can be written as:
\begin{widetext}
\bea \label{f of critical models}
f_z(N_0,m_+,m_-,\{m_j\};\{k_j\})=z^{N_0}g(z)(z-1)^{m_+}(z+1)^{m_-}\prod_{j=1}^{N_c}(z-e^{ik_j})^{m_j}(z-e^{-ik_j})^{m_j},
\eea
\end{widetext}
where $g(z)$ is a polynomial without any zeros on the unit circle or origin and $k_1<k_2<...<k_{N_c}$. Note that since  we have Hamiltonians with real couplings, all zeros are either real or come in complex conjugate pairs and all the powers are integers. For simplicity, we just consider $g(z)=1$. At this moment we assume that $N_0$ can be positive or negative integer number. In addition $m_+,m_-,m_j$ are non-negative integer numbers. The correlation matrix of the ground state for this model is shown to be\cite{Verresen:2019}:
\bea \label{G matrix}
G_{nm}=\f{1}{2\pi}\int_0^{2\pi}\f{f(e^{ik})}{|f(e^{ik})|}e^{-i(m-n)k}dk.
\eea
Remarkably the above integral can be calculated explicitly. The result for $g(z)=1$ can be written with respect to elementary functions as follows:
\begin{widetext}
\begin{equation}
G_{nm} = \left\{\begin{split}
&\f{4}{\pi}\f{\f{1}{2}G_{nm}^{Re}+G_{nm}^{Im}}{N_0+M+Q+n-m}& N_0+M+Q\neq m-n, \\
&(2\{\f{m_++1}{2}\}(-1)^{[\f{m_+}{2}]})((-1)^{M}+\f{4}{\pi}\sum_{j=1}^{N_c}(-1)^{j-1}\{\f{m_j}{2}\}k_j)& N_0+M+Q= m-n,\\ 
\end{split}\right.
\end{equation}
where we have

\begin{equation}
\left\{\begin{split}
&G_{nm}^{Re} = (-1)^{[\f{m_+}{2}]+1}(\{\f{m_+}{2}\}-\{\f{m_-}{2}\}(-1)^{[Q]+N_0}(-1)^{n+m})&\\
&G_{nm}^{Im} = \sum_{j=1}^{N_c}(-1)^{j-1}\{\f{m_j}{2}\}\sin(\f{\pi m_+}{2} + (N_0+M+Q+n-m)k_j)\\
\end{split}\right.
\end{equation}
and $M = \sum_{j=1}^{N_c} m_j $ , $Q = \f{m_++m_-}{2}$ and $\{X\}$ is defined as follows:
\begin{equation}
\{X\} := \mid X \mid - [\mid X \mid].
\end{equation}
\\
Using the above equation we find the following duality

\bea \label{G matrix duality I}
\textbf{G}[f_z(N_0,m_+,m_-,\{m_j\};\{k_j\})]=(-1)^{[\f{m_+}{2}]}
\textbf{G}[f_z(N_0+M[m_+,m_-,\{m_j\}],h[m_+],h[m_-],\{h[m_j]\};\{k_j\})],
\eea
\end{widetext}
where 
\bea \label{H function}
M[m_+,m_-,\{m_j\}]=[\f{m_+}{2}]+[\f{m_-}{2}]+2\sum_{j=1}^{N_c}[\f{m_j}{2}],\hspace{0.75cm}
\eea
and $h[x]=x-2[\f{x}{2}]$. For later use it is also useful to define
\bea \label{H function}
H[m_+,m_-,\{m_j\}]=h[m_+]+h[m_-]+2\sum_{j=1}^{N_c}h[m_j].\hspace{0.75cm}
\eea
 To the best of our knowledge the above duality has not been discussed before in the literature. It means that when $m_+,m_-,\{m_j\}$ are bigger than one it is possible to absorbe them to $N_0$ and remain with just one or zero powers for $m_+,m_-,\{m_j\}$. The immediate consequence of the above argument is that the even powers are non-critical and do not contribute to the central charge and the contribution of odd numbers is all the same. In other words we have the following theorem for the central charge:
\bea \label{central charge theorem}
c[N_0,m_+,m_-,\{m_j\};\{k_j\}]=\f{1}{2}H[m_+,m_-,\{m_j\}].\hspace{0.75cm}
\eea
Note that we assume that the models with $c=0$ are non-critical. In other words, all the models with $m_+,m_-,\{m_j\}$ even integer numbers are non-critical. From now on without loosing any generality we consider that $m_+,m_-,\{m_j\}$ are either zero or one and not all of them are zero. One can also prove another useful duality

\begin{widetext}
\bea \label{G matrix duality II}
\textbf{G}[f_z(N_0,m_+,m_-,\{m_j\};\{k_j\})]=-
\textbf{G}^T[f_z(-N_0-m_+-m_--2\sum_{j=1}^{N_c}m_j,m_+,m_-,\{m_j\};\{k_j\})].
\eea
\end{widetext}
Combining the two Equations \ref{G matrix duality I} and \ref{G matrix duality II} one can conclude that without loosing generality it is possible to assume that $m_+,m_-,\{m_j\}$ are either zero or one and $N_0$ is an integer number. To make the classification manageable and under control we will just consider the case $N_0=0$. 

\section{Summary of results}\label{sec:Summary of results}

In this section we will summarize our main results. We first discuss the case of the systems with $U(1)$ symmetry and then discuss the models without this symmetry. 

\subsection{Models with $U(1)$ symmetry}\label{subsec:Models with $U(1)$ symmetry}

Our extensive numerical results support the following behavior for the coefficient of the logarithm 
\be \label{xalpha U(1)}
x_{\alpha}=\left\{\begin{split}
&\f{c}{8}& \alpha\leq 4 \\
&\f{\alpha}{\alpha-1}\f{c}{8}& \alpha>4.\\ 
\end{split}\right.
\ee
The case of $c=1$ in a different geometry has been already discussed in \cite{Stephan:2014}. The presence of the discontinuity at $\alpha=4$ is attributed
to the least irrelevant operator in the Luttinger liquid description of the model. As far
as $\alpha<4$ it was argued in \cite{Stephan:2014} that this operator is
irrelevant and one can get $x_{\alpha}=\f{1}{8}$ by Luttinger
model arguments. However, when $\alpha>4$ this operator
is relevant and consequently the field gets locked into
one of the minima of the potential and just one of the configurations end up to have the largest contribution. Consequently, we have $x_{\alpha}=\f{\alpha}{\alpha-1}\f{c}{8}$. It seems this picture is more general and valid for generic $f_z(n)$ models. For $n=1$ a simple numerical investigation shows that the dominant configurations at $\alpha\to\infty$ are
$|0,1,0,1,...,0,1\rangle$ and $|1,0,1,0,...,1,0\rangle$ consistent with the half-filling ground state. It is possible to calculate the logarithm of the probability of this configuration exactly and one finds \cite{ARV:2021} a linear term plus a logarithmic subleading term with coefficient $-\frac{1}{8}$. This result proves the Eq \ref{xalpha U(1)} at $\alpha\to\infty$ for $n=1$. A simple numerical investigation shows that the largest probability for $f_z(n)$ models is attributed to the $2n$ configurations $|A_n\rangle=|\overbrace{0,0,...0}^{n-r}\overbrace{1,1,...1}^n,...,\overbrace{0,0,...0}^n\overbrace{1,1,...1}^r\rangle$ 
and $|A_n\rangle=|\overbrace{1,1,...1}^{n-r}\overbrace{0,0,...0}^n,...,\overbrace{1,1,...1}^n\overbrace{0,0,...0}^r\rangle$, where $r=0,1,...,n-1$. 
We conjecture that 
\be
-\ln p(A_n) = a(n) L+\f{n}{8}\ln L+\mathcal{O}(1).
\ee
It should be possible to prove the above conjecture using the methods developed in \cite{ARV:2021}, however, we do not attempt to do that in this paper. We note that when the system is not half filling similar picture is still valid but the most relevant configuration can change. For example, for $\f{r}{s}$ filling in $n=1$ case the most important configuration is $|A_1(\f{r}{s})\rangle=|\overbrace{0,0,...0}^{s-r}\overbrace{1,1,...1}^r,...,\overbrace{0,0,...0}^{s-r}\rangle$ in which the numerical results show that the logarithm of the probability of this configuration has also a linear term plus logarithmic correction with coefficient $-\frac{1}{8}$, see\cite{NR2016}.

Finally we also found that the Eq. \ref{xalpha U(1)} is most probably also valid for the models $f_z(\{a_n\})=\sum a_nf_z(n)$. The numerical results in these cases have strong oscillations and consequently the estimation for $x_{\alpha}$ is poor. However, the overall behavior of the numerical results is consistent with the Eq. \ref{xalpha U(1)}. 

The coefficient of the logarithm for all the considered models is summarized in the Table \ref{table:1}.

\begin{table*}[ht!]
\centering
\begin{tabular}{ |p{2cm}||p{2.5cm}|p{1.75cm}|p{2cm}|p{1.5cm}|  }
 \hline
 \multicolumn{5}{|c|}{Models with $U(1)$ symmetry} \\
 \hline
 \hspace{0.5cm}$f(z)$&\hspace{0.75cm} $f_z(1)$ &\hspace{0.5cm}$f_z(2)$&\hspace{0.25cm}$f_z(\{1,1\})$&$f_z(\{1,2\})$\\
 \hline
 \hspace{0.5cm}$8x_1$   & $0.9968\pm0.0001$    &$2.00\pm0.01$&   $1.003\pm0.003$&   $1.84\pm0.30$\\
  \hline
\end{tabular}
\caption{Coefficient of the logarithm in the Shannon entropy for different models with $U(1)$ symmetry.}
\label{table:1}
\end{table*}

\subsection{Models without $U(1)$ symmetry} \label{subsec:Models without $U(1)$ symmetry} 

For models with $N_0=0$ our numerical results done on many examples reveal the following behavior
\be \label{xalpha-no-U(1)}
x_{\alpha}=\left\{\begin{split}
&\f{b(\alpha)}{8}& \alpha\leq1 \\
&\f{\alpha}{\alpha-1}\f{c}{8}& \alpha>1\\ 
\end{split}\right.
\ee
where $b(\alpha)=\mathfrak{b}(\alpha) H[m_+,m_-,\{m_j\}]$ and $c=\f{1}{2}H[m_+,m_-,\{m_j\}]$. The coefficient of the logarithm seems to be again increasing based on the number of gapless Majorana fermions that one can define for the model. This is reminiscent of the behavior of entanglement entropy in these systems \cite{Its:2008}.
However, for $\alpha\leq1$ the coefficient is not exactly proportional to the central charge. For $\alpha=1$ we have $\mathfrak{b}(1)=0.480016\pm0.00005$ and for $0<\alpha<1$ the numerical results indicate a complicated but universal behavior, see \cite{Stephan:2014} for the Ising chain. There are also regions where this coefficient is negative. For all these models the most relevant configuration is the configuration without any fermion, i.e.  $|E\rangle=|0,0,...,0\rangle$ or the one with full of fermions $|E\rangle=|1,1,...,1\rangle$. One can understand this by calculating $\langle I| H|I\rangle$ for different configurations $I$. An easy calculation shows that $\langle I| H|I\rangle=nA_0$, where $n$ is the number of fermions in the configuration. It is now easy to see that depending on the sign of the $A_0$ just the configuration without any fermion or the one with full of fermions have the lowest energies. For the subsystem configurations numerical calculations support the above argument.
Note that just changing the sign of the $\bf{G}$ matrix interchanges the probability of the two configurations, however the set of the configurations is intact. We will be rarely concerned with this sign. The corresponding probability is called emptiness formation probability and one can calculate it explicitly using the Fisher-Hartwig formula, see \cite{Franchini:2005} for the $c=\frac{1}{2}$ case. In the most general case we find:
\be
-\ln p(E) = a(m_+,m_-,\{m_j\};\{k_j\}) L+\f{c}{8}\ln L+\mathcal{O}(1),
\ee
where $a(m_+,m_-,\{m_j\};\{k_j\})=\f{1}{2\pi}\int_{-\pi}^{\pi}\ln\f{1}{2}(1\mp\f{f(e^{ik})}{|f(e^{ik})|})$ and again we have $c=\f{1}{2}H[m_+,m_-,\{m_j\}]$.
This proves the Eq. \ref{xalpha-no-U(1)} for $\alpha\to\infty$. However, as it is argued already for Ising chain in \cite{Stephan:2014} it is not clear why the discontinuity in $x_{\alpha}$ should start exactly at $\alpha=1$ in all of these models. 

In the Table \ref{Table:2} we summarize the coefficient of the logarithm for all the models where we did comprehensive numerical checks.
\begin{table*}[ht!]
\centering
\begin{tabular}{ |p{2cm}||p{3cm}|p{3cm}|p{2.35cm}|p{3cm}|  }
 \hline
 \multicolumn{5}{|c|}{Models without $U(1)$ symmetry} \\
 \hline
 \hspace{0.5cm}$f(z)$&\hspace{1cm} $z-1$ &\hspace{0.8cm} $(z-1)_{BW}$&\hspace{0.55cm}$z^2-1$&\hspace{1cm}$z^3-1$\\
 \hline
 \hspace{0.5cm}$8x_1$   & $0.48008\pm0.00002$   &$0.48009\pm0.00003$&  $0.9616\pm0.0001$& \hspace{0.25cm}   $1.4465\pm0.0008$\\
  \hline
\end{tabular}
\caption{Coefficient of the logarithm in the Shannon entropy for different models without $U(1)$ symmetry.}
\label{Table:2}
\end{table*}


\section{Numerical and fitting procedure}\label{sec:Numerical and fitting procedure} 

In this section we briefly discuss our numerical and fitting procedures. The more comprehensive details are relegated to the Appendix \ref{App:fitting}.

In all of the considered models we first find the $2^L$ number of probabilities using the Eq. (\ref{formation probability}). The largest size that we considered was $L_{max}=42$. After collecting all the probabilities we calculate the $I_{\alpha}$ and find the best estimate of $x_{\alpha}$ using different fitting  procedures. Most importantly our fitting function is 

\be \label{fitting function}
I=A_0+A_1\log L+A_2L^{-1}\log L+\sum_{i=3}^mA_i/L^{i-2}.
\ee
However, there are at least two important challenges to overcome. First of all, due to the limitation in the maximum size of $L$ we need to use some extrapolation methods to get a good estimate of $x_{\alpha}$. The second important hurdle is that $I_{\alpha}$ for some of the models show strong oscillations. In these cases either one needs to stick to a particular branch or average the estimated $x_{\alpha}$  over all the branches. The more sophisticated approach is to use the regularization method. We have tried all of these possibilities and in each case we report the one with the best fit possible. In the Appendix \ref{App:fitting} we also explain in detail our methods to estimate the error bars in each case.

Because of the exponential nature of the calculations and the number of considered models computing all the probabilities required a quite long time, which is particularly notable for larger system sizes. As an example in the case of $L=42$, it took about $3$ days to generate $2^{42}$ formation probabilities using a cluster with $356$ computing nodes, where each node had $16$ cores. To prevent further damage to the environment in Appendix \ref{App:Shannon} we collected the Shannon entropy for the models that we considered so that the motivated reader can reproduce the coefficient of the logarithm by her(him)self.

\section{Details of the analysis}\label{sec:Details of the analysis} 

In this section we will provide the details of the models that we considered.
We first discuss systems with $U(1)$ symmetry and later we discuss the ones without this symmetry.

\subsection{Models with $U(1)$ symmetry} \label{subsec:Details of the analysis-Models with U(1) symmetry} 

\begin{figure}[htp]
  \centering
  \includegraphics[width=0.5\textwidth]{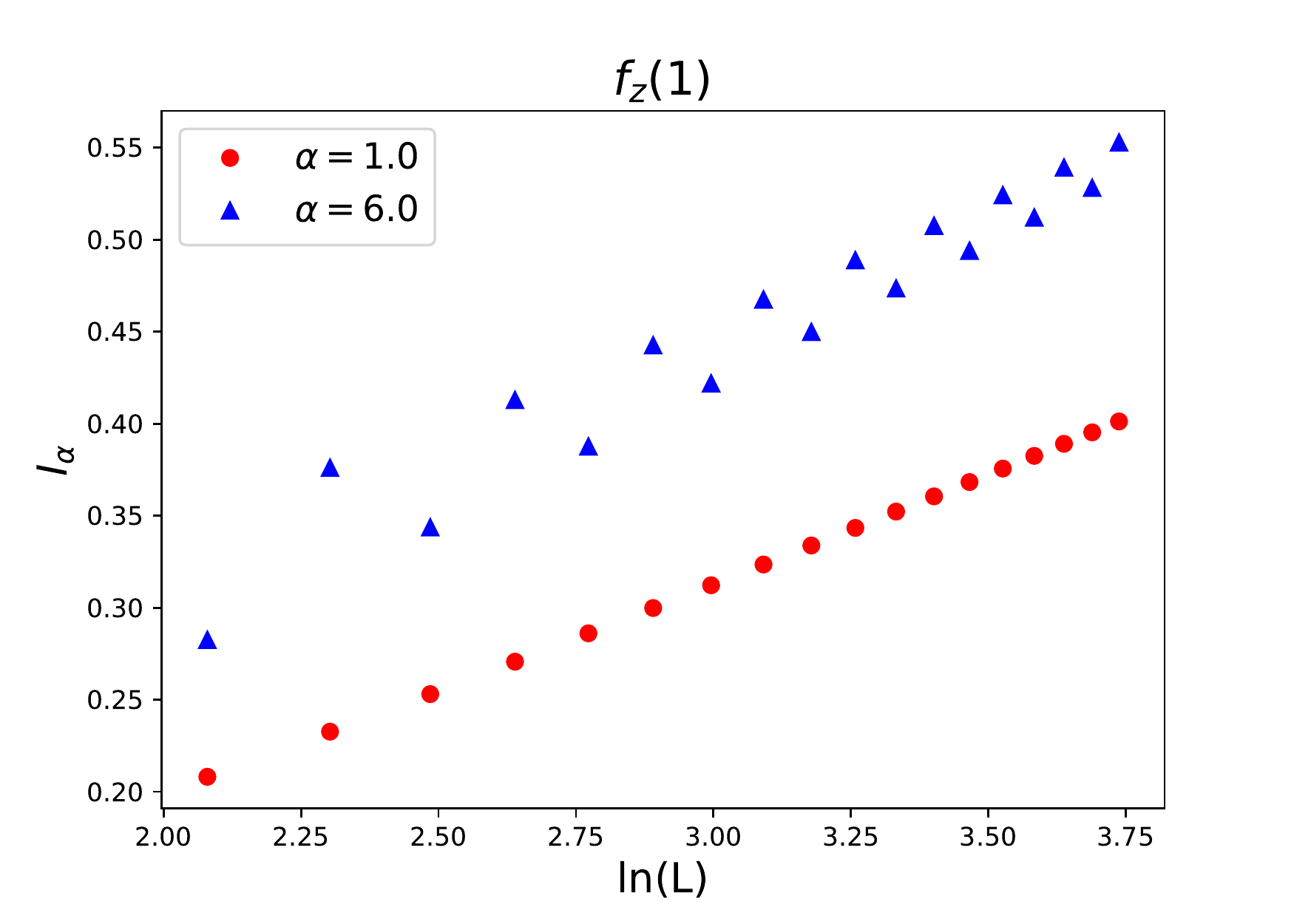}\\
  \caption{$I_{\alpha}$ with respect to $\ln L$ for $f_z(1)$ for two indices $\alpha=1$ and $\alpha=6$. }
  \label{f_z(1)}
\end{figure}

\begin{figure}[ht]
  \centering
  \includegraphics[width=0.5\textwidth]{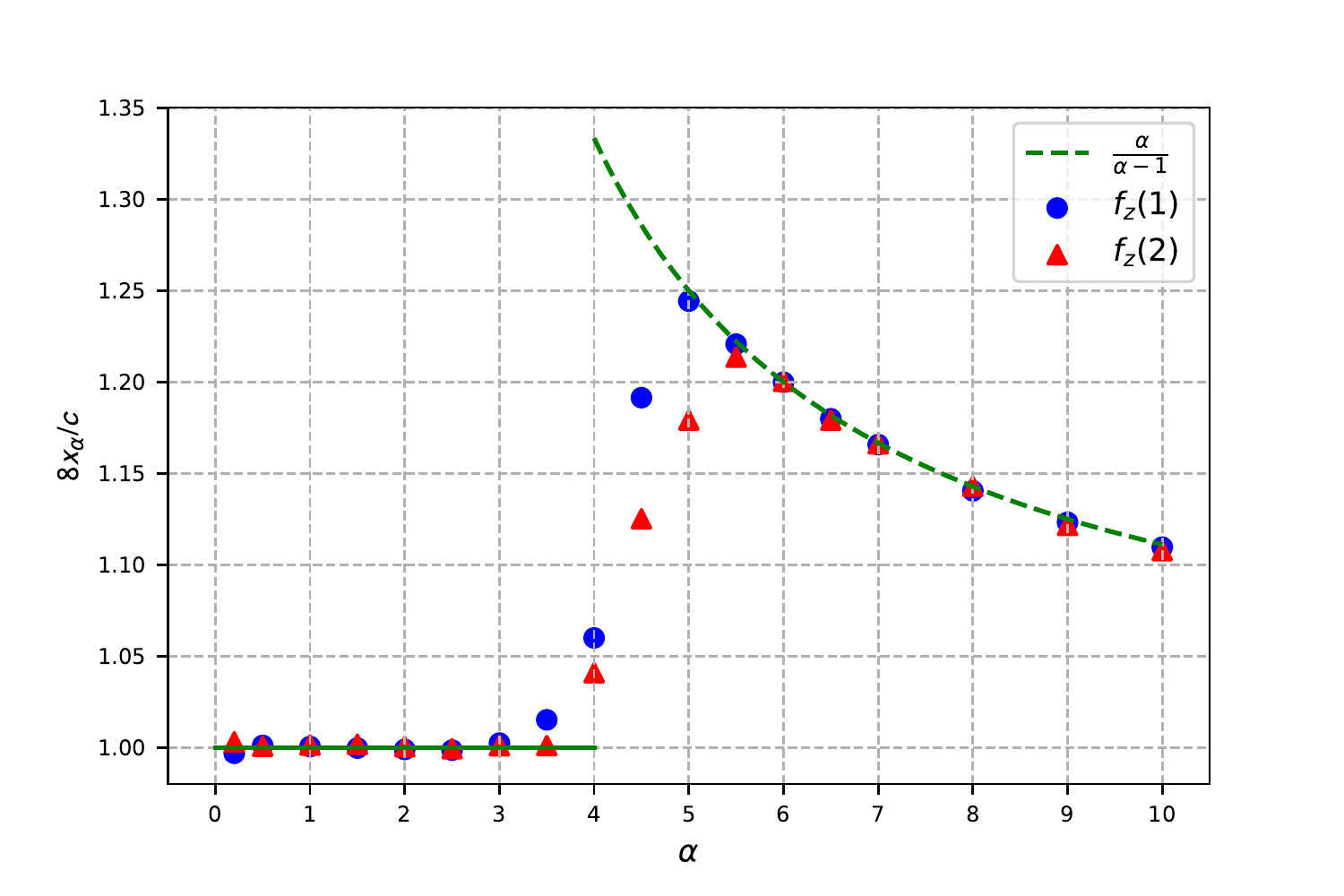}\\
  \caption{The coefficient of the logarithm with respect to $\alpha$ for the two models $f_z(1)$ and $f_z(2)$. In all the calculations $L_{max}=42$. }
  \label{coefficent_U1}
\end{figure}

\begin{figure}[ht]
  \centering
  \includegraphics[width=0.5\textwidth]{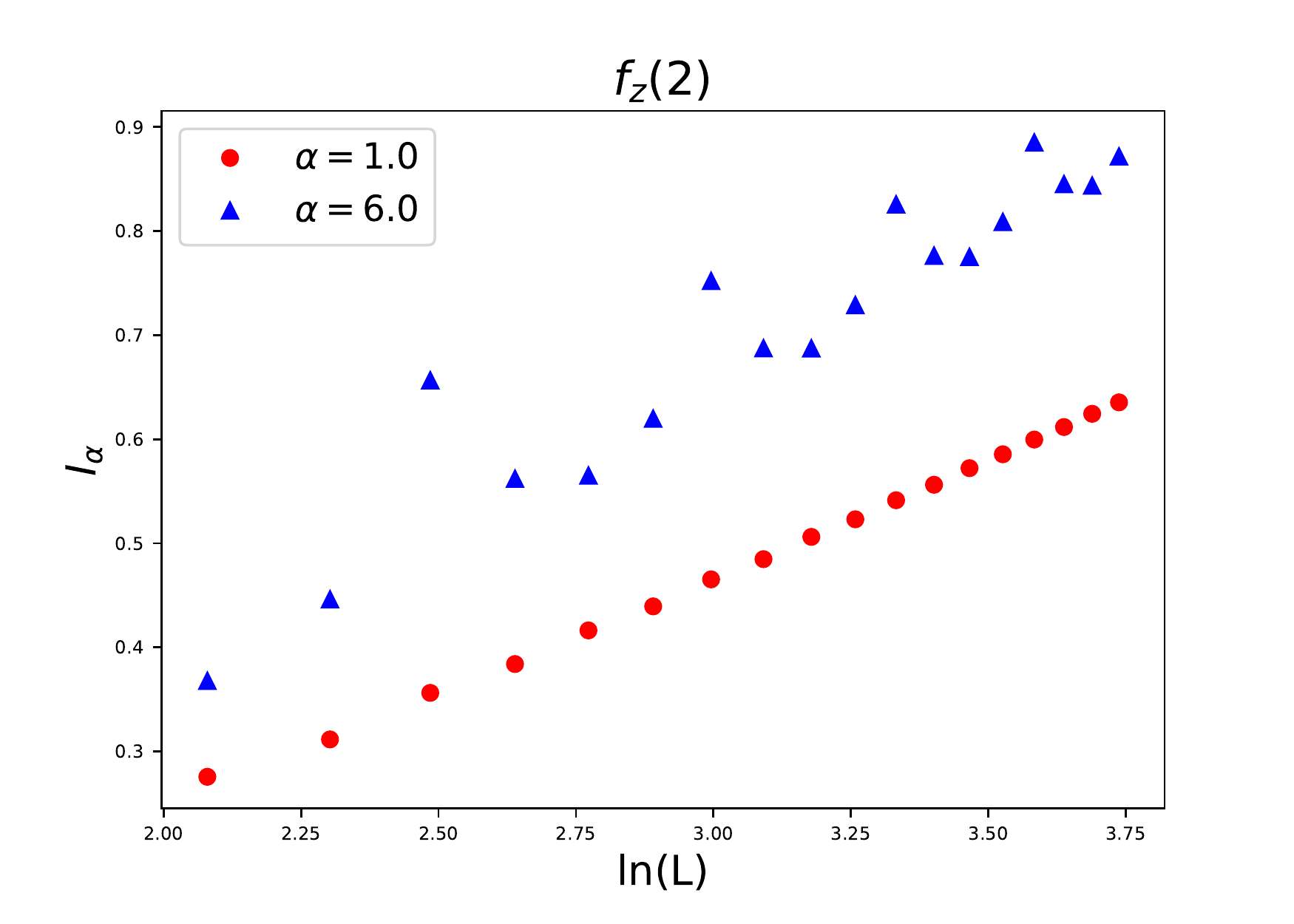}\\
  \caption{$I_{\alpha}$ with respect to $\ln L$ for $f_z(2)$ for two indices $\alpha=1$ and $\alpha=6$. }
  \label{f_z(2)}
\end{figure}

We first considered the model $f_z(1)$ which is the simple hopping model with half filling. The results for $I_{\alpha}$  with $\alpha=1,6$ are shown in Fig \ref{f_z(1)}. The results for $\alpha>1$ have oscillations which gets stronger by increasing $\alpha$. To calculate the coefficient of the logarithm in these cases we first calculated the coefficient for each branch using extrapolation method and later we averaged over the two results. The results are shown in the Fig \ref{coefficent_U1} which is compatible with the Eq. \ref{xalpha U(1)}.
We then considered the model $f_z(2)$. The results for $I_{\alpha}$  with $\alpha=1,6$ are shown in Fig \ref{f_z(2)}. There are stronger oscillations in this case.
There are four visible branches for $\alpha>1$. In these cases again we calculated the coefficient for each branch and if needed we also used the regularization method as it is explained in the Appendix \ref{App:fitting}. Finally we averaged over all the branches.
The coefficient $x_{\alpha}$ with respect to $\alpha$ is shown again in the Fig \ref{coefficent_U1}.

We also considered the models $f_z(\{a_1,a_2\})$ with $(a_1,a_2)\in\{(1,1),(1,2)\}$. The numerical results have strong oscillations especially for the case $(a_1,a_2)=\{(1,2)\}$. In this case for large $\alpha$'s it seems impossible to get a good estimate for the $x_{\alpha}$ with the sizes up to $L=42$. However, the general picture is consistent with the Eq. \ref{xalpha U(1)}. In the Appendix \ref{App:Shannon} we just report the results for the Shannon entropy and do not show the details for the other $\alpha$'s.

\subsection{Models without $U(1)$ symmetry} \label{subsec:Details of the analysis-Models without U(1) symmetry} 

In this section we will provide some details regarding the models without 
$U(1)$ symmetry. 

\begin{figure}[ht]
  \centering
  \includegraphics[width=0.5\textwidth]{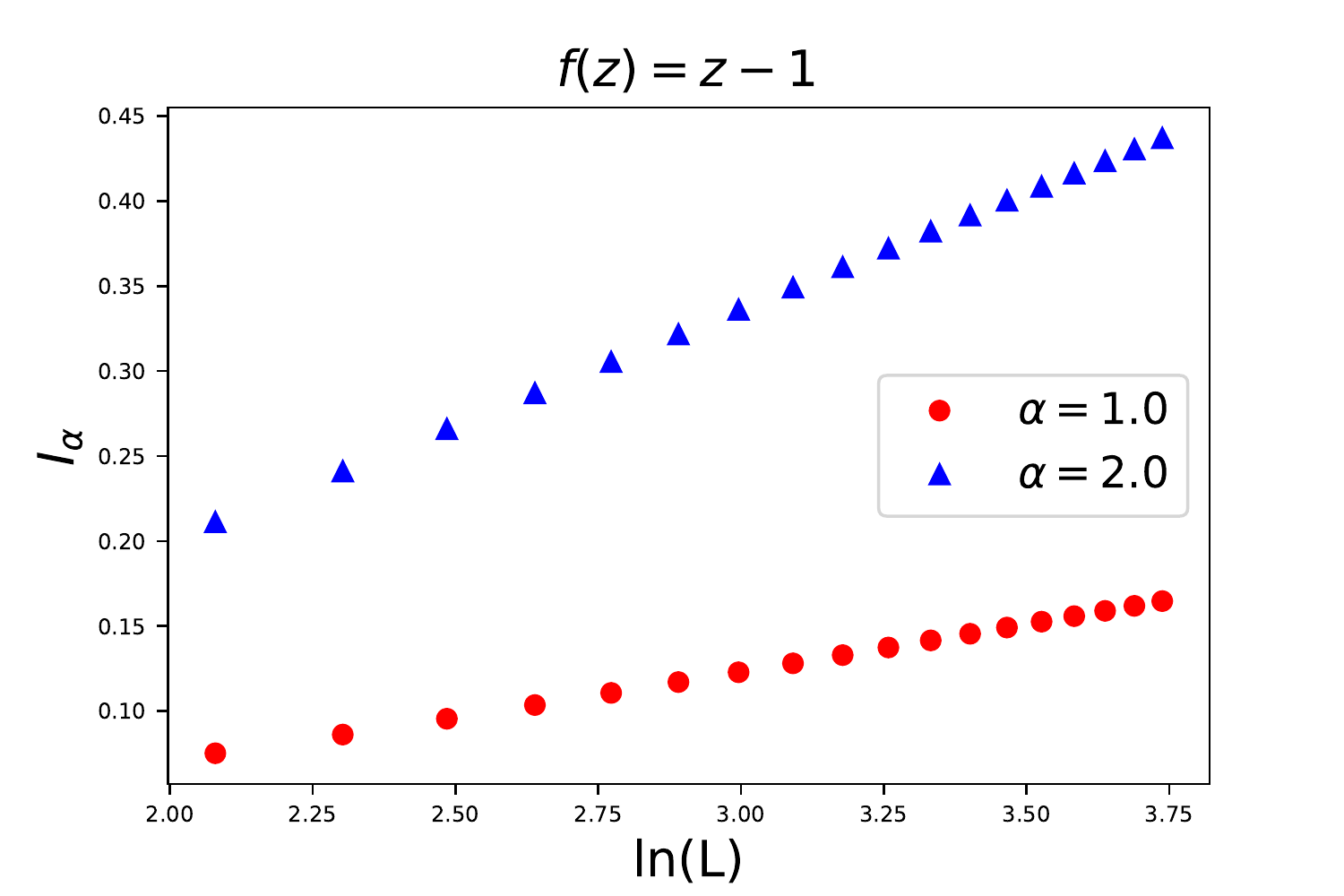}\\
  \caption{$I_{\alpha}$ with respect to $\ln L$ for $f(z)=z-1$ for two indices $\alpha=1$ and $\alpha=2$. }
  \label{f(z-1)}
\end{figure}
\begin{figure}[ht]
  \centering
  \includegraphics[width=0.5\textwidth]{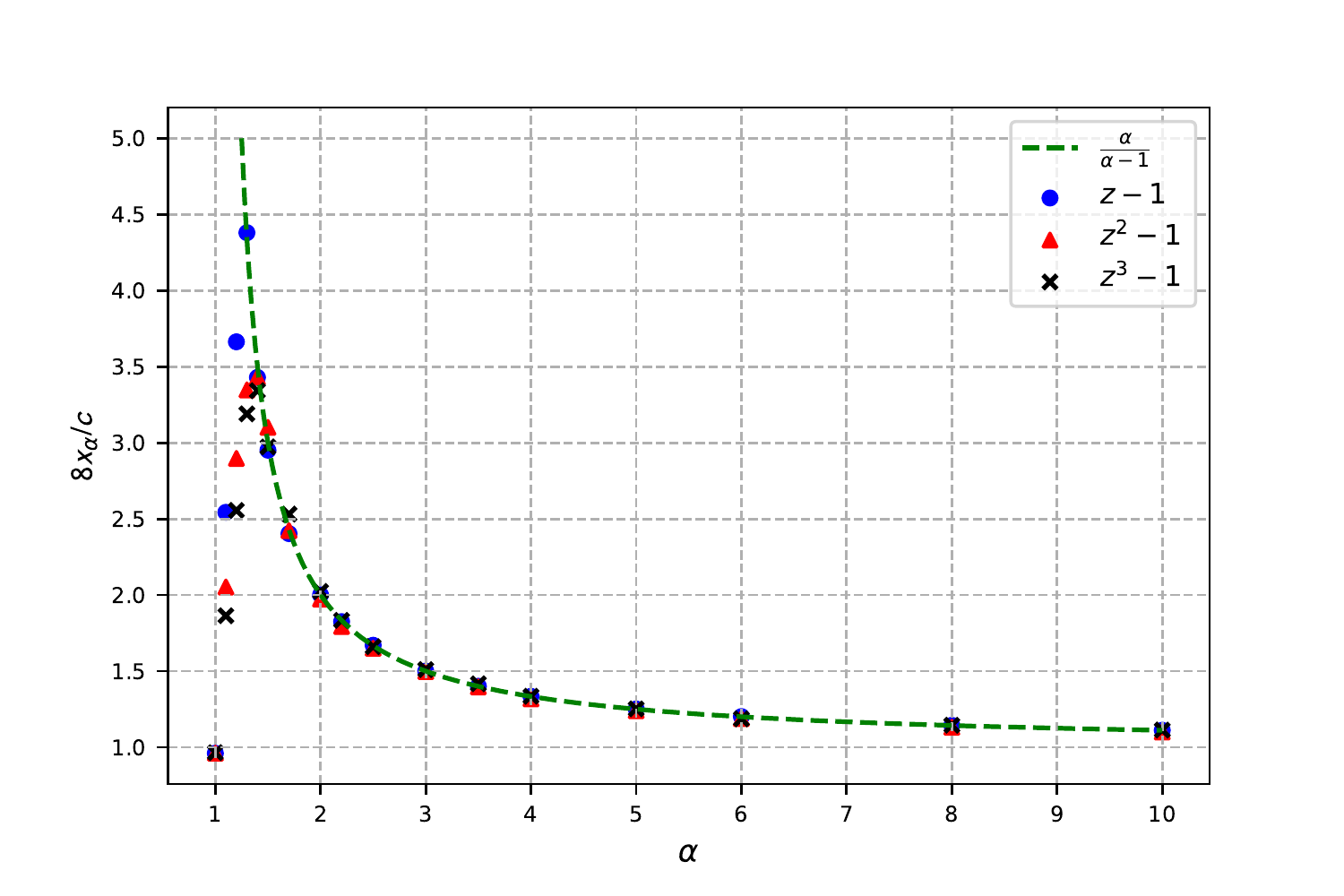}\\
  \caption{The coefficient of the logarithm with respect to $\alpha$ for the three models $f(z)=z-1$, $f(z)=z^2-1$ and $f(z)=z^3-1$. In all the calculations $L_{max}=42$. }
  \label{coefficent_non_U1}
\end{figure}

The first example is the famous Ising chain with $f(z)=z-1$. The results for $I_{\alpha}$  with $\alpha=1,2$ are shown in Fig \ref{f(z-1)}. We do not see any oscillations for any $\alpha$. To calculate the coefficient of the logarithm we used the extrapolation method explained in the Appendix \ref{App:fitting}. The maximum size of the subsystem that we considered was $L_{max}=42$.
The results for $\alpha\geq1$ are shown in the Fig \ref{coefficent_non_U1} which is consistent with the Eq. \ref{xalpha-no-U(1)}.

It is worth mentioning that we also analyzed the $f(z)=z+1$ which although has different $\textbf{G}$ matrix the set of formation probabilities are exactly the same as the Ising chain.

\begin{figure}[ht]
  \centering
  \includegraphics[width=0.5\textwidth]{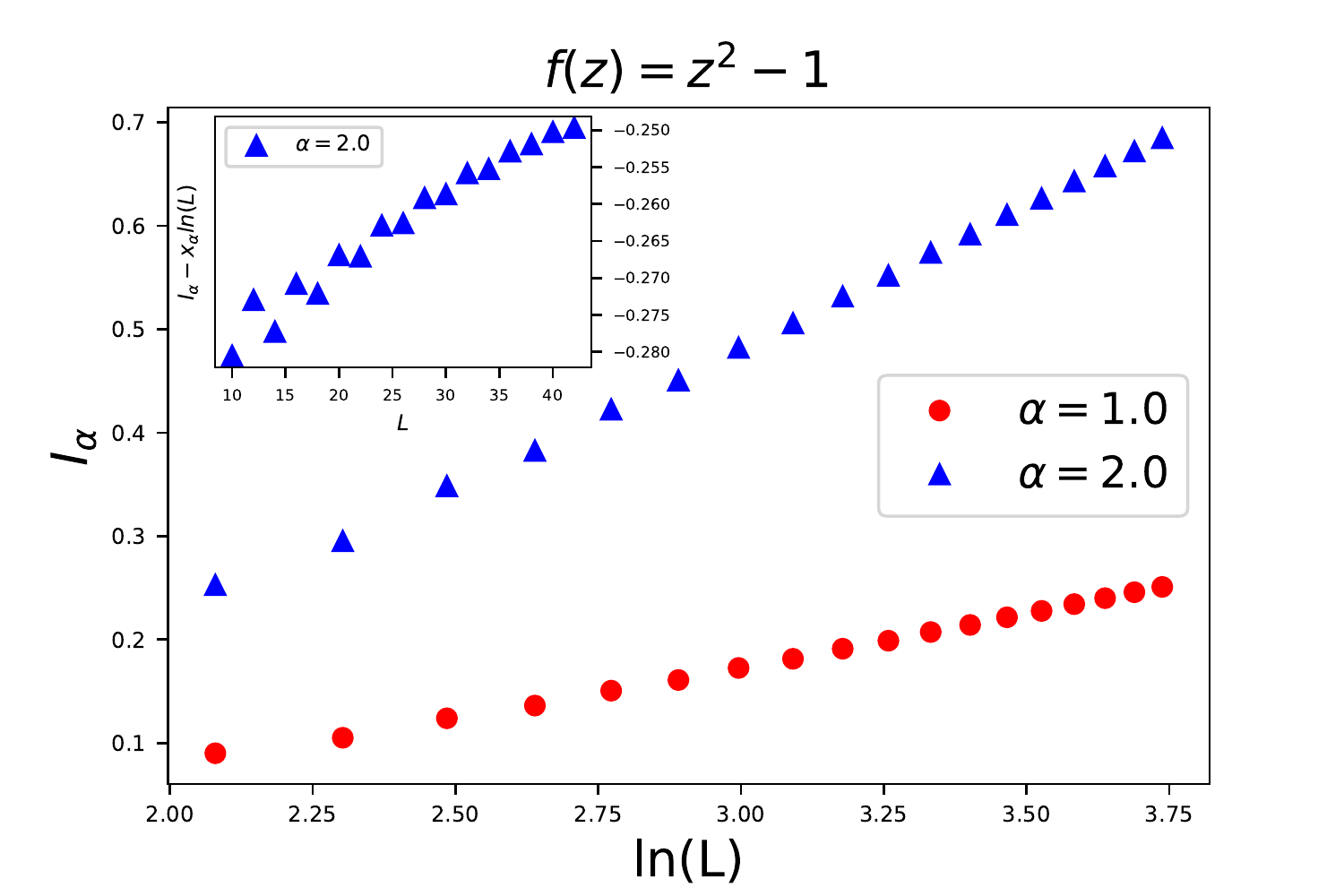}\\
  \caption{$I_{\alpha}$ with respect to $\ln L$ for $f(z)=z^2-1$ for two indices $\alpha=1$ and $\alpha=2$. The inset shows there are oscillations with period two.}
  \label{f(z-1)(z+1)}
\end{figure}

The second example is  $f(z)=z^2-1$ which is a model with central charge $c=1$. The results for $I_{\alpha}$  with $\alpha=1,2$ are shown in Fig \ref{f(z-1)(z+1)}. There are small oscillations for $\alpha>1$ which are just detectable after careful numerical manipulations, see Fig \ref{f(z-1)(z+1)} inset. To calculate the coefficient of the logarithm we again separated different branches and used the extrapolation method for each branch and then finally averaged over the two branches. The results for $\alpha\geq1$ are shown in the Fig \ref{coefficent_non_U1} which is consistent with the Eq. \ref{xalpha-no-U(1)}. Note that although the central charge here is an integer number because of lack of $U(1)$ symmetry we end up to a result which resembles the one we obtained for the Ising chain. 

We also analyzed other models such as $f(z)=(z-e^{i\theta})(z-e^{-i\theta})$ with different $\theta$'s. They all have $c=1$ and show similar structure. We realized that when $\theta$ is small or close to $\pi$ the oscillations for $\alpha>1$ are stronger. The fewest oscillations appear for $\theta=\f{\pi}{2}$  which have the same set of probabilities as $f(z)=z^2-1$.

\begin{figure}[ht]
  \centering
  \includegraphics[width=0.5\textwidth]{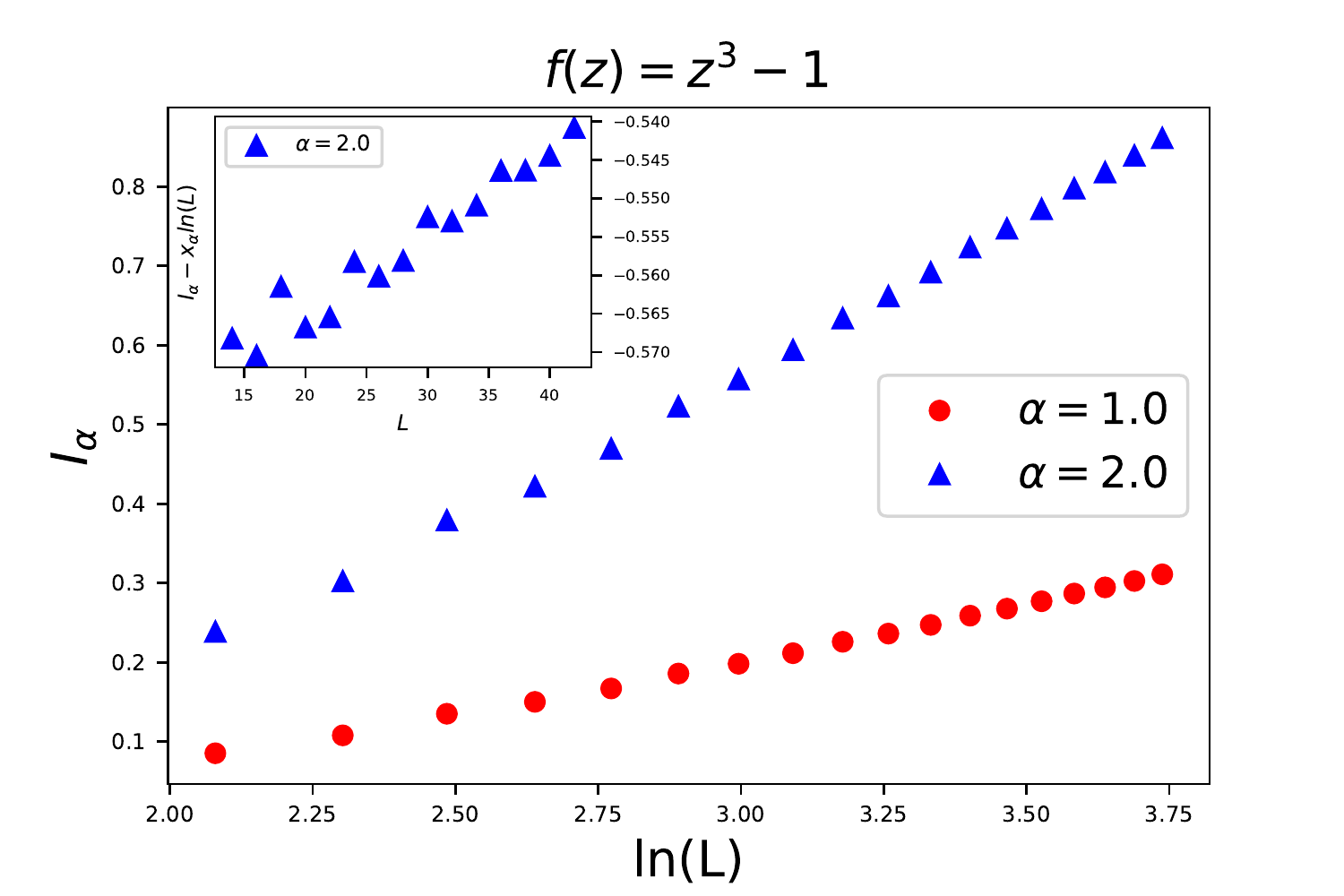}\\
  \caption{$I_{\alpha}$ with respect to $\ln L$ for $f(z)=z^3-1$ for two indices $\alpha=1$ and $\alpha=2$. The inset shows there are oscillations with period three.}
  \label{fc32}
\end{figure}

The third example is $f(z)=z^3-1$ which is a model with central charge $c=\f{3}{2}$. The results for $I_{\alpha}$  with $\alpha=1,2$ are shown in Fig \ref{fc32}. Similar to the previous case we have small oscillations. There are three branches and we followed the same procedure as before to estimate the coefficient of the logarithm. The results for $\alpha\geq1$ are shown in the Fig \ref{coefficent_non_U1} which is again consistent with the Eq. \ref{xalpha-no-U(1)}. 

We also considered other models with similar central charge such as $f(z)=(z-1)(z-e^{i\theta})(z-e^{-i\theta})$ with different $\theta$'s. The result are the same as before. However, we realized that the case $\theta=\f{2\pi}{3}i$ has the fewest oscillations. When we decrease or increase $\theta$ the oscillations get stronger. Similar phenomena happens also for $f(z)=(z+1)(z-e^{i\theta})(z-e^{-i\theta})$. When the zeros has the largest distance from each other the oscillations are smallest and when two or three of them get closer to each other we have stronger oscillations. This is a numerical observation which we do not have a good explanation.

Apart from the above case we also considered $f(z)=(z-1)(z+1)(z-e^{i\theta})(z-e^{-i\theta})$ with central charge $c=2$ with again similar conclusions. $f(z)=z^4-1$ has the least oscillations. The last model we considered was $f(z)=z^5-1$ with the central charge $c=\f{5}{2}$. The results are consistent with the Eq. \ref{xalpha-no-U(1)}. In most of the cases where we do not report the results here we considered $L_{max}=36$. In some cases such as $f(z)=z^5-1$ we pushed the results up to $L_{max}=42$.

\begin{figure}[ht]
  \centering
  \includegraphics[width=0.5\textwidth]{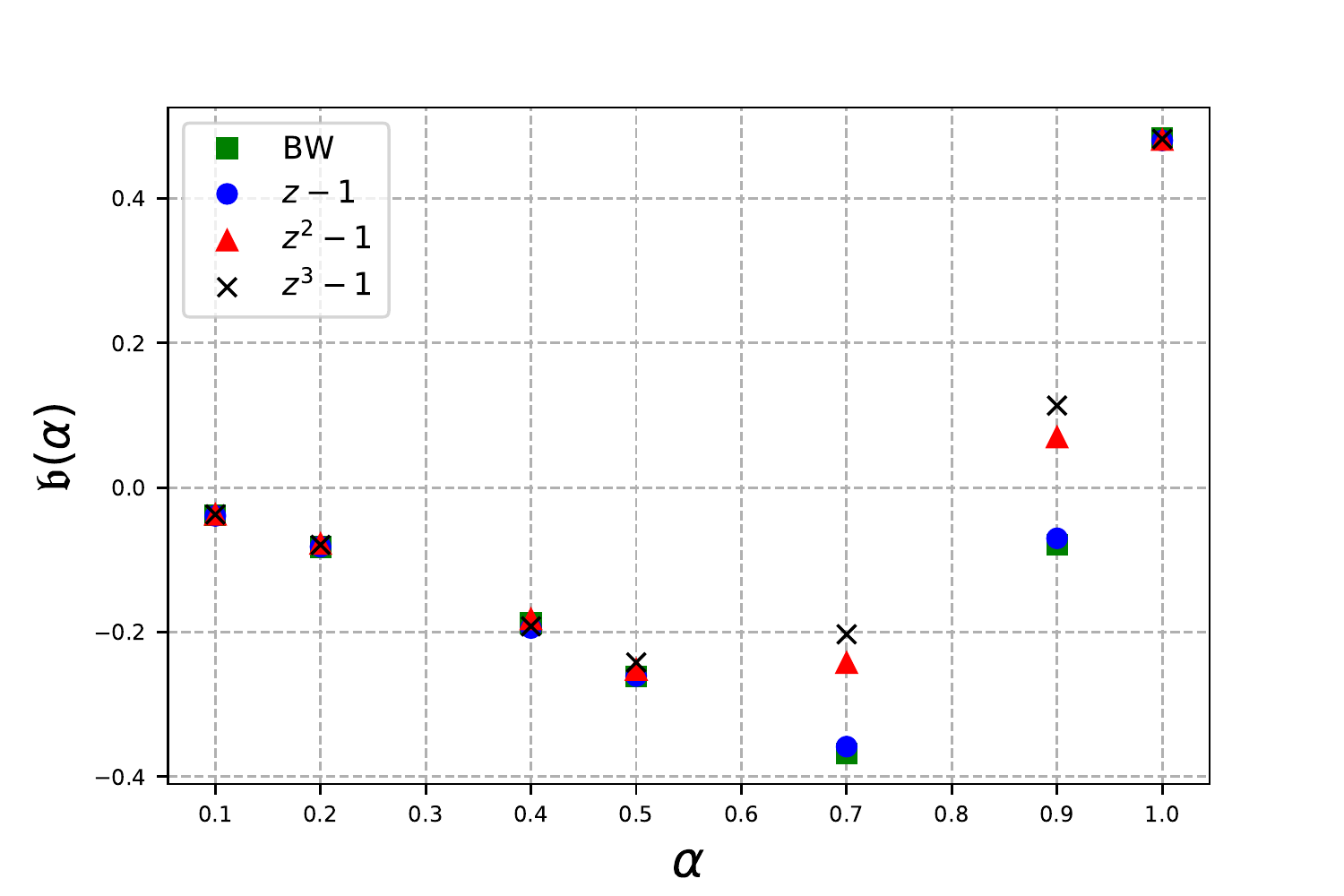}\\
  \caption{The coefficient of the logarithm with respect to $\alpha$ for the three models $f(z)=z-1$, $f(z)=z^2-1$ and $f(z)=z^3-1$. In all the calculations $L_{max}=36$. }
  \label{0-alpha-1}
\end{figure}

In all of the above cases we also studied with the same procedure $I_{\alpha}$ with $\alpha<1$. For some $\alpha$'s the coefficient of the logarithm is negative but the numerical results confirm that the behavior of the coefficient is universal and proportional to the number of points where one can linearize the dispersion relation. The results were depicted in Fig \ref{0-alpha-1}. Here we considered $L_{max}=36$. We think the visible discrepancy in the region $\alpha\in(0.6,1)$ is due to the finite size effect which for unknown reason to us is stronger in this interval.

\section{Analysis based on Bisognano-Wichmann reduced density matrix}
\label{sec:BW} 

The reduced density matrix (RDM) of a quantum system, $\r_A$, is fully encoded in the modular (or entanglement) Hamiltonian $H_A$ defined as
\be
\r_A = \f{\ep^{- H_A}}{Z_A}, \qquad Z_A = \tr_A \ep^{- H_A}.
\ee
By construction, the RDM and the modular Hamiltonian have the same eigenvectors, and their eigenvalues are simply related. The modular Hamiltonian plays a key role in quantum field theory~\cite{Witten:2018lha}.
In this context, the modular Hamiltonian of half-space partition is known to be related to the boost operator~\cite{Bisognano:1975ih,Bisognano:1976za}. Its form in conformal field theory (CFT) is also known explicitly
\cite{Hislop:1981uh,Najafi:2016kwb}. However, its explicit functional form in lattice models is known only in a few simple 
cases, see for example
\cite{itoyama1987lattice,peschel1999density,nienhuis2009entanglement,peschel2009reduced,eisler2017analytical,eisler2018properties}.
It was proposed in Refs.~\cite{Dalmonte:2017bzm,Giudici:2018izb} to use the Bisognano-Wichmann (BW) theorem
in quantum field theory and its extension in conformal field theory (CFT)
 to write approximate modular Hamiltonians for lattice models.
From the BW modular Hamiltonian one can construct a RDM, which has been dubbed BW RDM. The proposal has been checked extensively
\cite{Dalmonte:2017bzm,Giudici:2018izb,eisler2018properties,Mendes-Santos:2019ine,Mendes-Santos:2019tmf,Zhang:2020mjv},
showing that in many cases the BW modular Hamiltonian can reproduce to a good precision the entanglement spectrum, correlation functions and entanglement entropy. In \cite{Zhang:2020mjv} it was shown that this approximation also produces very good approximations of the formation probabilities. For a recent comprehensive review see \cite{Dalmonte:2022}

Since the BW modular Hamiltonian is a discretization of the quantum field theory itself one might hope that the convergence of many quantities to the actual field theory result might be faster and better. Having this in mind we used the BW of the Ising model \cite{eisler2018properties} to find first the $\bf{G}_{BW}$ matrix as follows: we first make the following $T$ matrix:
\begin{eqnarray}\label{Ising exact G}\
e^{\left(
\begin{array}{cc}
 \bf{M} & \bf{N} \\
 -\bf{N} & -\bf{M}
\end{array}
\right)}=\left(
\begin{array}{cc}
 \bf{T}_{11} & \bf{T}_{12} \\
 \bf{T}_{21} & \bf{T}_{22}
\end{array}
\right),
\end{eqnarray} 
where the $\bf{M}$ and $\bf{N}$ are $L\times L$ matrices ($L$ is the size of subsystem) with the following elements:
\begin{eqnarray}\label{M and N}\
\frac{1}{\pi}N_{lm}&=&\lambda(l)\delta_{l+1,m}-\lambda(m)\delta_{l,m+1},\\
\frac{1}{\pi}M_{l,m}&=&\lambda(l)\delta_{l+1,m}+\lambda(m)\delta_{l,m+1}+2\lambda(l-\frac{1}{2})\delta_{l,m},\hspace{0.5cm}
\end{eqnarray}
where $\lambda(n)=\frac{n(L-n)}{L}$ and $L$ is the size of the subsystem, i.e. $L=1,2,3,...$.
Then for  $\bf{G}$ matrix we have 
\begin{eqnarray}\label{GBW}\
\bf{G}_{BW}=\frac{\bf{F}-I}{\bf{F}+I},
\end{eqnarray}
where $\bf{F}=\bf{T}_{22}^{-1}+\bf{T}_{12}.\bf{T}_{22}^{-1}$. One can use the above correlation matrix to produce formation probabilities and consequently the R\'enyi (Shannon) entropy. To calculate $I_{\alpha}(l)$ one needs to take care of a subtlety. The BW reduced density matrix is not an exact reduced density matrix. That means  $\tr_{\bar{B}}\rho_A^{BW}\ne\rho_B^{BW}$. In other words, probabilities coming from $\tr_{\bar{B}}\rho_A^{BW}$ and $\rho_B^{BW}$ are different. We realized that the best results for $I_{\alpha}(l)$ can be derived by using marginal probabilities of $\rho_L^{BW}$.
The results for the R\'enyi entropy is indistinguishable from the exact results when depicted in the Figure so we just report the Shannon entropy in this case in the Appendix \ref{App:Shannon}.
This is an interesting demonstration of the universality of the coefficient of the logarithm and also the power of the approximate BW modular Hamiltonian.

\section{Discussion}\label{sec:Discussion} 

In this paper we considered an infinite system and calculated the coefficient of the logarithm appearing in the scaling of R\'enyi (Shannon) entropy of the ground state of critical chains. Instead of working directly in the thermodynamic limit one could take a finite periodic system with size $N$ and calculate $I_{\alpha}=Re_{\alpha}(\ell)+Re_{\alpha}(N-\ell)-Re_{\alpha}(N)$ which ends up to be proportional to $x'_{\alpha}\ln\f{N}{\pi}\sin\f{\pi \ell}{N}$, see \cite{AR:2013,Stephan:2014,AR:2014}. We expect that in all of our models $x'_{\alpha}=x_{\alpha}$. The same is not true if one takes an open boundary condition as it was already noticed in \cite{Stephan:2014}. This is because the boundary conditions can change the logarithmic subleading term drastically. Clear understanding of the coefficient in open quantum spin chains is still lacking. All of the models that we considered in this paper can be mapped to quantum spin chains using Jordan-Wigner transformation. Based on the previous numerical calculations \cite{AR:2013,AR:2014} it seems plausible to assume that if one calculates the R\'enyi (Shannon) entropy in $\sigma^x$ basis the result for the coefficient of the logarithm be the same as what we found in this paper. However, as it was already noticed in \cite{AR:2014} this might not be correct for other bases. Finally we should mention that understanding the coefficient of the logarithm for $\alpha\leq1$ in critical systems without $U(1)$ symmetry appears to be a challenge. For $\alpha=1$ this coefficient seems almost \cite{AR:2013} but not exactly \cite{Stephan:2014} proportional to the central charge. It is an open problem to understand why this is the case.

\section{Conclusions}\label{sec:Conclusions} 

R\'enyi (Shannon) entropy of the ground state of quantum chains shows a volume-law behavior. When the system is critical these quantities for the subsystem show a subleading logarithmic term with a coefficient which is universal up to some extent. In this paper we studied these quantities at the critical point of generic time-reversal translational invariant quadratic critical free fermions. We found that there are two different classes of models. Models with $U(1)$ symmetry show a unified behavior. The coefficient is dependent on the number of points one can linearize the dispersion relation. The coefficient is constant up to $\alpha=4$ and then there is a discontinuity and a nice decay in the form $\f{\alpha}{\alpha-1}$. In the case of systems without $U(1)$ symmetry we have studied models where the corresponding $f(z)$ function has no zero outside of the unit circle. In these models the coefficient of the logarithm is always proportional to the number of points where one can linearize the dispersion relation. There is a discontinuity at $\alpha=1$ and for $\alpha>1$ we again have the $\f{\alpha}{\alpha-1}$ kind of decay. For $\alpha\leq1$ although the coefficient is still universal the exact $\alpha$ functionality is not known. There are also regions where this coefficient is negative. 
It would be interesting to generalize the above analysis to models in which $g(z)\neq1$ and/or $N_0\neq0$. Due to numerous possibilities
and the existence of strong oscillations in the calculation of the R\'enyi entropy the complete classification in these cases might not be strightforward.
Finally we also studied the same quantities using the approximate BW modular Hamiltonian and confirmed that the produced set of formation probabilities are very close to the exact ones. The derived coefficient of the logarithm was almost indistinguishable from the exact results.
The biggest challenge for the future is probably to calculate analytically the $\mathfrak{b}(\alpha)$ for the second class of models to understand the nature of these numbers.

\section*{Acknowledgements}\label{sec:Acknowledgements}

We thank K. Najafi for early collaboration on the subject.
MAR thanks CNPq and FAPERJ (Grant No. 210.354/2018) for partial support.
\appendix

\section{Details of the fitting procedure}\label{App:fitting}

In this section we provide more details regarding fitting procedures that we followed in the main text. Let $\left\lbrace \left(L_j,I_j \right) \right\rbrace_{j=1}^n$ be the set of data points, in which we intend to extract the relevant physical quantities, such as the coefficient of the logarithm, as the fitting parameters. The fitting functions of interest in this paper are of a factorized form meaning that it is a direct sum of some fitting (known) functions where the fitting parameters are the corresponding pre-factors. In other words, we have
\begin{equation}\label{eq:hypothesis}
h(L,\left\lbrace \theta\right\rbrace )=\sum_{i=0}^{m}\theta_if_i(L),
\end{equation}
where $m$ is the number of the fitting terms. In the above $f_i(L)$, $i=0,...,m$ are the fitting functions assumed to be \textit{a priori} known ($f_0(L)\equiv 1$) and $\left\lbrace \theta\right\rbrace$ shows the set of the fitting parameters, i.e. $\left\lbrace \theta_0,\theta_1,...,\theta_m\right\rbrace$. In particular, $\theta_0$ is called the bias. To make connection with the paper, where the fitting formula is $I=A_0+A_1\log L+A_2 L^{-1}\log L+\sum_{i=3}^mA_i/L^{i-2}$, we have $\theta_0\equiv A_0$, $(\theta_1,f_1(L))=(A_1\equiv x_{\alpha},\log L)$, $(\theta_2,f_2(L))=(A_2,L^{-1}\log L)$, and $(\theta_{i},f_i(L))=(A_i,L^{-(i-2)})$, $i=3,...,m$. The $\chi^2$ method is an efficient approach for estimating the best fitting for a given data set, see for example \cite{Stephen Boyd}. 
One defines the $\chi^2$ as follows
\begin{equation}\label{eq:chi_sq}
\chi^2(\left\lbrace \theta\right\rbrace ) = \frac{1}{n}\sum_{j=1}^n\left[I_j-h(L_j,\left\lbrace \theta\right\rbrace ) \right]^2,
\end{equation} 
which should be minimized with respect to all the fitting parameters $\left\lbrace \theta\right\rbrace$ in order to get the best fitting. In our work we mostly worked with another quantity called $R$ value. It is defined as $R^2\equiv 1-\frac{\chi^2}{\sigma^2}$, where $\sigma^2\equiv \frac{1}{n}\sum_{j=1}^n\left(I_j-\bar{I} \right)^2 $ and $\bar{I}\equiv\frac{1}{n}\sum_jI_j$. The closer this quantity is to one the better the fit is.
When the number of fitting parameters are high, one can use the gradient descent method in which one  updates the parameters using the equation
\begin{equation}
\vec{\theta}_i^{\text{new}}=\vec{\theta}_i^{\text{old}}- \eta\frac{\nabla_{\theta} \chi(L_j,\left\lbrace \theta\right\rbrace )}{\left| \nabla_{\theta} \chi(L_j,\left\lbrace \theta\right\rbrace )\right| }\delta \theta.
\end{equation}
where $\delta \theta$ is a discretization parameter,  $\eta$ is the step size and
$\vec{\theta}\equiv \left(\theta_0,\theta_1,...,\theta_m \right) $, and $\nabla_{\theta}\equiv \left(\frac{\partial}{\partial\theta_0},\frac{\partial}{\partial\theta_1},...,\frac{\partial}{\partial\theta_m} \right)$ to get the best fit after reaching to the fixed point of the parameter.
It is worth metioning that one should be careful in taking appropriate number of fitting parameters to avoid the problem of \textit{over-fitting}. Normally the sign of overfitting in numerical calculations is the huge and strongly fluctuating numbers for the fitting parameters.
 To avoid this problem, one adds $\lambda\sum_{i=2}^m\theta_i^2$ to the $\chi^2$, where $\lambda$ is a very small coefficient which prevents the coefficients to take extremely large values. This method is called the \textit{regularization method} and was used in this paper when necessary. Note that in this work we did not regularize the $\theta_0$ and $\theta_1$.

A more compact representation of Eq.~\ref{eq:chi_sq} can be obtained by casting the equation in a matrix form. Let $\textbf{X}_{ji}\equiv f_i(L_j)$ be a $(n)\times (m+1)$ matrix, and $\Theta\equiv (\theta_0,\theta_1,...,\theta_m)^T$ be a vector with length $m+1$, and $Y\equiv(I_1,I_2,...,I_n)^T$ where $i=0,1,...,m$ enumerates the fitting terms, and $j=1,2,...,n$ enumerates the data points. Then the regularized $\chi^2$ reads 
\begin{equation}
\chi^2(\left\lbrace \theta\right\rbrace )=\frac{1}{n}\left(\textbf{X}\Theta-Y \right)^T\left(\textbf{X}\Theta-Y \right)+ \lambda\Theta^T\textbf{I}'\Theta,
\end{equation}
where $\textbf{I}'$ is a diagonal matrix with zero or one as its diagonal elements. For a fitting parameter that is not going to be regularized, the corresponding diagonal element is zero, and for the other elements, it is one. By minimizing $\chi^2$ with respect to all $\theta$ parameters, we obtain
\begin{equation}
\Theta =\left( \textbf{X}^T\textbf{X}+\lambda\textbf{I}'\right)^{-1}\textbf{X}^TY.
\end{equation}
In some cases the matrix $\textbf{X}^T\textbf{X}$ has small eigenvalues which leads to very large $\theta$ values when $\lambda$ is zero. This is the reason for introducing the regularization parameter $\lambda$.

In our analyses in this paper, we set the diagonal elements of $\textbf{I}'$ corresponding to the bias and  the $x_{\alpha}$ to zero. In addition, when overfit takes place, we consider a minimal $\lambda$ value that removes the over-fit. To obtain an optimal values for $m$, we first start with $m=1$, and find the fittings. Then we increase $m$ by one and repeat the fittings, and check the convergence of the fitting parameters. We continue this procedure, comparing the quality of the fittings with the previous stage, up to a stage where the fittings are optimal. \\
To calculate the fitting parameters in the scaling limit we used two kinds of extrapolation: the uppermost and the lowermost fixed extrapolation (UFE and LFE respectively). Suppose that the range of fitting is $[L_{\text{min}},L_{\text{max}}]$. Then in the UFE (LFE) method we fix $L_{\text{max}}$ ($L_{\text{min}}$) to its maximum (minimum) value and calculate the fitting parameters (especially $x_{\alpha}$) in terms of $L_{\text{min}}$ ($L_{\text{max}}$) using the $R^2$ method. Our observations show that for all cases the resulting  $x_{\alpha}(L)$ follow
\begin{equation}
x_{\alpha}(L)=x_{\alpha}^{\infty}+\frac{a}{L^b},
\label{Eq:FittingPowerLaw}
\end{equation}
where $x_{\alpha}^{\infty}$, $a$ and $b$ are some constants obtained by fitting. $x_{\alpha}^{\infty}$ is the extrapolated parameter that we report in this paper. We note that the UFE is not really the usual extrapolation method because the largest size is actually fixed. However, in most of the cases since the values for the R\'enyi entropy for small sizes are not very useful we found that the UFE normally gives more stable results than the LFE.

Throughout the paper we face models where the R\'enyi entropy shows some oscillations. In this cases we subdivide the data points to $k$ classes where in each class the points are in the same phase of oscillations: $\left\lbrace \left\lbrace (L_i^{(q)},I_i^{(q)})\right\rbrace_{i=1}^n\right\rbrace_{q=1}^k $. Then, using the procedure explained above we find the best fits \textit{for each class}, with the resulting fitting parameters $\left\lbrace \left\lbrace \theta_{p}^{(q)}\right\rbrace_{p=0}^m\right\rbrace_{q=1}^k $, where $q$ numerates the classes (totally $k$ classes). Then, the average parameters are simply defined as
\begin{equation}
\bar{\theta}_p\equiv \frac{1}{k}\sum_{q=1}^k\theta_{p}^{(q)},
\end{equation}
for which the corresponding fitting functions are free of oscillations. The mentioned oscillations are stronger in $U(1)$ symmetric models. \\

Furthermore, one can define error of estimating coefficients of fitting, using standard deviation method (SD) \cite{Harvey Gould}. We know that for any set of data with linear behavior we have a finite deviations (errors) from Eq.    \ref{eq:hypothesis}. For $n$ data pairs $\{(L_i, I_i), i = 1, ..., n\}$, the underlying relationship between $I_i$ and $L_i$ including this error term $\epsilon_i$ can be described as
\begin{equation}\label{eq:error}
I(L_j) = A_{-1} + x_{\alpha} \ln L_j+... + \epsilon_j.
\end{equation}
After applying proper fitting method and extracting fitting coefficients we can replace them in Eq. \ref{eq:error} and estimate error $\epsilon_j=I_i-A_{-1}-x_{\alpha} \ln L_j-...$ . Consequently, the error of estimating coefficient $x_{\alpha}$ can be written as
\begin{equation}
\epsilon({x_{\alpha}})=\sqrt{\frac{\sum_j\epsilon_j^2}{(n-2)(<z^2> - <z>^2)}},
\end{equation} 
where $z=\ln L$ and $<a> = \frac{1}{n}\sum_{j}a_j$.\\

\begin{figure}[h]
	\centering
	\includegraphics[width=0.60\linewidth]{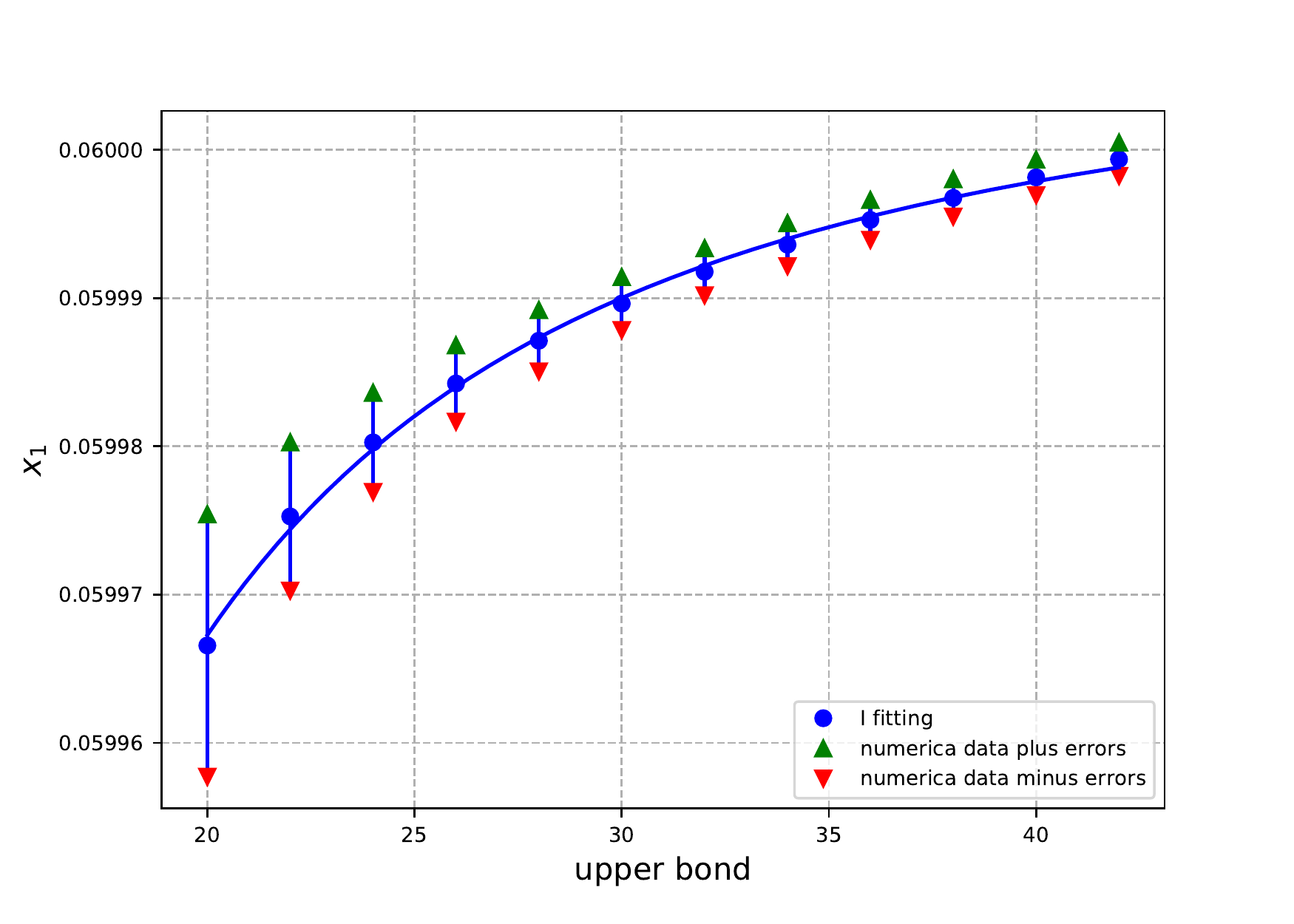}
	\caption{Blue points are $x_{\alpha}$ estimations for each size derived using Eq.~\ref{eq:hypothesis}.
	Green/Red points are $x_{\alpha}$ plus/minus the error bar of the blue points.	
	To estimate the error bar for $x_\alpha$, the difference between $x_{\alpha}^{\infty}$ associated to the blue, red and green points are used.}
	\label{fig:error_explain}
\end{figure}

There are two sources of error  in estimating the thermodynamic limit of fitting parameters, like $x_{\alpha}$. The first one is the systematic error corresponding to Eq.~\ref{Eq:FittingPowerLaw}. The second one is the error arising from the numerical errors in estimating the function for each system size. See for example Fig.\ref{fig:error_explain}, where apart from the systematic error in estimating the final value of $I$ for large $L_{\text{max}}$ values, there is an additional error due to the errors for each $L_{max}$. In other words, we use the Eq.~\ref{Eq:FittingPowerLaw} for three sets of points, the fitting values of $x_{\alpha}$ coming from Eq.~\ref{eq:hypothesis} for different sizes and the same numbers plus/minus their error bars for each sizes.
More precisely, in Fig.\ref{fig:error_explain} although the blue bold circles converge to $0.06001052$ with error $1.2 \times 10^{-6}$, the upward (downward) triangles converge to $0.06001193$ ($0.06000933$) as a limit of maximum (minimum) estimations. The total error is then the summation of these two error bars. In the example of Fig.\ref{fig:error_explain}, the total error is $\pm \left[0.06001193-0.06001052 +1.2\times 10^{-6} \right] $. To summarize here we first calculate the extrapolation value of the green, blue and red points. Then we calculate the difference between the values associated to red and green points with the blue one. Then we pick the maximum value and add it to the error bar of the fitting of the blue points.

\section{List of Shannon entropies}\label{App:Shannon}
In this appendix we summarize the Shannon entropy for different sizes for all the models that we considered in this paper.

\subsection{Shannon entropy of models with $U(1)$ symmetry}
In this section we provide the exact values of Shannon entropy in the models with $U(1)$ symmetry.
 In table \ref{table:shannon-U1} we provided the Shannon entropy for different sizes for the models we considered in the main part of the paper.

	\begin{table*}[ht!]
		\centering
		\begin{tabular}{ |p{0.8cm}||p{3cm}|p{3cm}|p{3cm}|p{3cm}| }
			\hline
			\multicolumn{5}{|c|}{Models with $U(1)$ symmetry} \\
			\hline
			L&\hspace{1cm} $f_z(1)$ &\hspace{0.5cm}$f_z(2)$&$f_z(1,1)$&$f_z(1,2)$\\
			\hline
			$1$   & $0.69314718055994$    &$0.69314718055994$ & $0.63651416829481$&
			$0.69221884672917$\\
			\hline
			$2$   & $1.30175595835581$    &$1.38629436111989$ & $1.20713991833245$&
			$1.38401888518543$\\
			\hline
			$3$   & $1.88952473643207$    &$1.99490313891575$& $1.75382909171979$&
			$2.01139838075075$\\
			\hline
			$4$   & $2.46572926830832$    &$2.60351191671162$& $2.29024160464855$&
			$2.63699346158892$\\
			\hline
			$5$   & $3.03554130502831$    &$3.19128069478788$& $2.82016064054523$&
			$3.25084607202052$\\
			\hline
			$6$   & $3.60094114646480$    &$3.77904947286415$& $3.34574544378459$&
			$3.86112826700399$\\
			\hline
			$7$   & $4.16329822887889$    &$4.35525400474039$& $3.86834033023438$& $4.46758265294401$\\
			\hline
			$8$   & $4.72333514587764$    &$4.93145853661664$& $4.38865348976192$&
			$5.07065792051820$\\
			\hline
			$9$   & $5.28160174014818$    &$5.50127057333664$& $4.90719393994777$&
			$5.67191793745351$\\
			\hline
			$10$   & $5.83843850171108$   &$6.07108261005664$& $5.42433643749537$&
			$6.27020185844070$\\
			\hline
			$11$   & $6.39411934070096$    &$6.63648245149308$& $5.94032544493927$&
			$6.86734966048696$\\
			\hline
			$12$   & $6.94883095743287$    &$7.20188229292957$& $6.45535416818048$&
			$7.46264647774241$\\
			\hline
			$13$   & $7.50272914860883$    &$7.76423937534372$& $6.96957656772588$&
			$8.05706979765335$\\
			\hline
			$14$   & $8.05592711783444$    &$8.32659645775783$& $7.48310440162153$&
			$8.65038954334950$\\
			\hline
			$15$   & $8.60852182009970$    &$8.88663337475661$& $7.99603073579124$&
			$9.24285216104092$\\
			\hline
			$16$   & $9.16058699436123$    &$9.44667029175474$& $8.50843325729873$&
			$9.83467192662273$\\
			\hline
			$17$   & $9.71218703560899$    &$10.0049368860264$& $9.02037209149640$&
			$10.4256806854124$\\
			\hline
			$18$   & $10.2633726246811$    &$10.5632034802959$& $9.53189899717722$&
			$11.0162682615565$\\
			\hline
			$19$   & $10.8141886934818$    &$11.1200402418602$& $10.0430585231179$&
			$11.6061362664382$\\
			\hline
			$20$   & $11.3646715586883$    &$11.6768770034228$& $10.5538866526276$&
			$12.1956618677633$\\
			\hline
			$21$   & $11.9148538081393$    &$12.2325578424122$& $11.0644150816116$&
			$12.7846447900688$\\
			\hline
			$22$   & $12.4647623481134$    &$12.7882386814019$& $11.5746716896468$&
			$13.3732928226244$\\
			\hline
			$24$   & $13.5638519339168$    &$13.8976619148654$& $12.5944598914257$&
			$14.5494962592860$\\
			\hline
			$26$   & $14.6620982637734$    &$15.0054582972159$& $13.6134084894882$&
			$15.7244854175891$\\
			\hline
			$28$   & $15.7596246309596$    &$16.1118542356559$& $14.6316397420975$&
			$16.8984152066857$\\
			\hline
			$30$   & $16.8565291237941$    &$17.2170436401148$& $15.6492512752802$&
			$18.0714355315951$\\
			\hline
			$32$   & $17.9528910588046$    &$18.3211739881711$& $16.6663221275571$&
			$19.2436688684954$\\
			\hline
			$34$   & $19.0487754831082$    &$19.4243740680976$& $17.6829169075352$& $20.4152009971075$\\
			\hline
			$36$   & $20.1442363964986$    &$20.5267452314426$& $18.6990893971895$& $21.5861034172132$\\
			\hline
			$38$   & $21.2393191176783$    &$21.6283773787716$& $19.7148848398415$&
			$22.7564395029841$\\
			\hline
			$40$   & $22.3340608777584$    &$22.7293430908899$& $20.7303414022332$ & $23.9262777420356$\\
			\hline
			$42$   & $23.4284982110165$    &$23.8297075943315$& $21.7454918339490$ & $25.0956690768337$\\
			\hline
		\end{tabular}
		\caption{Shannon entropy for different sizes for various models with $U(1)$ symmetry.}
		\label{table:shannon-U1}
	\end{table*}

\subsection{Shannon entropy of models without $U(1)$ symmetry}
In this section we provide the exact values of Shannon entropy in the models without $U(1)$ symmetry. In table \ref{table:shannon-no-U1-I} we provided the Shannon entropy for different sizes for the models with $f(z)=\{z-1,z^2-1,z^3-1,z^5-1,z(z-1)\}$. Then in the table \ref{table:shannon-BW} first column we provided the Shannon entropy coming from the BW reduced density matrix. In the second column we explicitly write the Shannon entropy for the subsystem derived using the mariginal probabilities.

\begin{table*}[ht!]
		\centering
		\begin{tabular}{ |p{0.8cm}||p{3cm}|p{3cm}|p{3cm}|p{3cm}| }
			\hline

			\hline
			L&\hspace{1cm} $z-1$ &$z^2-1$&$z^3-1$&$z^5-1$\\
			\hline
			$1$   & $0.47394663373377$    &$0.47394663373377$& $0.47394663373377$&$0.47394663373377$ \\
			\hline
			$2$   & $0.92544105529219$    &$0.94789326746755$& $0.94789326746755$& $0.94789326746755$\\
			\hline
			$3$   & $1.36797061201631$    &$1.39938768902597$& $1.42183990120133$& $1.42183990120133$\\
			\hline
			$4$   & $1.80585459307135$    &$1.85088211058439$& $1.87333432275975$& $1.89578653493511$\\
			\hline
			$5$   & $2.24088987072848$    &$2.29341166730851$& $2.32482874431817$ & $2.36973316866888$\\
			\hline
			$6$   & $2.67400379724519$    &$2.73594122403263$& $2.77632316587659$& $2.82122759022730$\\
			\hline
			$7$   & $3.10573474075415$    &$3.17382520508767$& $3.21885272260071$& $3.27272201178572$\\
			\hline
			$8$   & $3.53642296390859$    &$3.61170918614271$& $3.66138227932483$& $3.72421643334414$\\
			\hline
			$9$   & $3.96629704662543$    &$4.04674446379984$& $4.10391183604894$& $4.17571085490256$\\
			\hline
			$10$   & $4.39551790695338$    &$4.48177974145697$& $4.54179581710399$& $4.62720527646098$\\
			\hline
			$11$   & $4.82420308419465$    &$4.91489366797366$& $4.97967979815903$& $5.06973483318511$\\
			\hline
			$12$   & $5.25244103454533$    &$5.34800759449037$& $5.41756377921407$& $5.51226438990921$\\
			\hline
			$13$   & $5.68029998893319$    &$5.77973853799942$& $5.85259905687121$& $5.95479394663334$\\
			\hline
			$14$   & $6.10783367902435$    &$6.21146948150821$& $6.28763433452831$& $6.39732350335750$\\
			\hline
			$15$   & $6.53508517170340$    &$6.64215770466274$& $6.72266961218539$& $6.83985306008173$\\
			\hline
			$16$   & $6.96208951510668$    &$7.07284592781729$& $7.15578353870211$& $7.27773704113673$\\
			\hline
			$17$   & $7.38887561225378$    &$7.50272001053396$& $7.58889746521889$& $7.71562102219158$\\
			\hline
			$18$   & $7.81546757783182$    &$7.93259409325053$& $8.02201139173568$& $8.15350500324665$\\
			\hline
			$19$   & $8.24188574022752$    &$8.36181495357858$& $8.45374233524467$& $8.59138898430154$\\
			\hline
			$20$   & $8.66814739454094$    &$8.79103581390823$& $8.88547327875337$& $9.02927296535642$\\
			\hline
			$21$   & $9.09426737732450$    &$9.21972099114686$& $9.31720422226235$& $9.46430824301328$\\
			\hline
			$22$   & $9.52025851138501$    &$9.64840616838662$& $9.74789244541657$& $9.89934352067105$\\
			\hline
			$24$   & $10.3718974749895$    &$10.5048820690910$& $10.6092688917255$& $10.7694140759866$\\
			\hline
			$26$   & $11.2231381653332$    &$11.3605999778672$& $11.4690170571602$& $11.6375632801578$\\
			\hline
			$28$   & $12.0740383772891$    &$12.2156673580469$& $12.3281120002033$& $12.5037911331902$\\
			\hline
			$30$   & $12.9246441747402$    &$13.0701703434076$& $13.1865537208623$& $13.3700189862121$\\
			\hline
			$32$   & $13.7749928645435$    &$13.9241790301996$& $14.0439240753413$ & $14.2334808732707$\\
			\hline
			$34$   & $14.6251150843050$    &$14.7777512244745$& $14.9008472029367$& $15.0969427602233$\\
			\hline
			$36$   & $15.4750363025723$    &$15.6309351556417$& $15.7573231036387$& $15.9593619268318$\\
			\hline
			$38$   & $16.3247779202053$    &$16.4786149708125$& $16.6130410124114$& $15.8455641809791$\\
			\hline
			$40$   & $17.1743580924403$    &$17.3362947859834$& $17.4684336568878$& $16.6467809808711$\\
			\hline
			$42$   & $18.0237923539904$    &$18.1885347499710$& $18.3235010370748$& $18.5387292548529$\\
			\hline
		\end{tabular}
		\caption{Shannon entropy for different sizes for various models without $U(1)$ symmetry.}
		\label{table:shannon-no-U1-I}
	\end{table*}

	\begin{table}[ht!]
	\centering
	\begin{tabular}{ |p{1cm}||p{3cm}|p{3cm}|  }
		\hline
		\multicolumn{3}{|c|}{BW} \\
		\hline
		L&\hspace{0.5
		cm} total system &\hspace{0.5cm}subsystem\\
		\hline
		$1$   & $0.45920500717509$    &$-$\\
		\hline
		$2$   & $0.92205905719825$    &$0.47270562810661$\\
		\hline
		$3$   & $1.36645271889283$    &$-$\\
		\hline
		$4$   & $1.80496685397933$    &$0.92510102900855$\\
		\hline
		$5$   & $2.24030399857181$    &$-$\\
		\hline
		$6$   & $2.67358767640636$    &$1.36780423258969$\\
		\hline
		$7$   & $3.10542377417024$    &$-$\\
		\hline
		$8$   & $3.53618169812639$    &$1.80575515259345$\\
		\hline
		$9$   & $3.96610437840695$    &$-$\\
		\hline
		$10$   & $4.39536048183681$    &$2.24082354163218$\\
		\hline
		$11$   & $4.82407203544144$    &$-$\\
		\hline
		$12$   & $5.25233024242084$    &$2.67395633633817$\\
		\hline
		$13$   & $5.68020509249127$    &$-$\\
		\hline
		$14$   & $6.10775148615426$    &$3.10569907398650$\\
		\hline
		$15$   & $6.53501329186441$    &$-$\\
		\hline
		$16$   & $6.96202612243598$    &$3.53639516977199$\\
		\hline
		$17$   & $7.38881928804390$    &$-$\\
		\hline
		$18$   & $7.81541720297786$    &$3.96627477249654$\\
		\hline
		$19$   & $8.24184042007064$    &$-$\\
		\hline
		$20$   & $8.66810640526907$    &$4.39549965444791$\\
		\hline
		$21$   & $9.09423012703542$    &$-$\\
		\hline
		$22$   & $9.52022451133198$    &$4.82418785310530$\\
		\hline
		$24$   & $10.3718688193031$    &$5.25242813140267$\\
		\hline
		$26$   & $11.2231136875357$    &$5.68028891764912$\\
		\hline
		$28$   & $12.0740172271869$    &$6.10782407525548$\\
		\hline
		$30$   & $12.9246257180375$    &$6.53507676179803$\\
		\hline
		$32$   & $13.7749766184651$    &$6.96208208951522$\\
		\hline
		$34$   & $14.6251006747952$    &$7.38886900778397$\\
		\hline
		$36$   & $15.4750234355686$    &$7.81546166549282$\\
		\hline
		$38$   & $16.3247663610014$    &$8.24188041670934$\\
		\hline
		$40$   & $17.1743476517357$    &$8.66814257612601$\\
		\hline
		\hline
	\end{tabular}
	\caption{Shannon entropy for different sizes for Ising chain derived using BW reduced density matrix. The third column is the marginal probabilities for the half of the subsystem.}
	\label{table:shannon-BW}
	\end{table}
\hfill\\
\newpage

\providecommand{\href}[2]{#2}\begingroup\raggedright\endgroup

\end{document}